\documentclass[amsmath,amssymb,aps, prb,twocolumn, superscriptaddress, showkeys]{revtex4-2}
\usepackage{graphicx} % Required for inserting images
\usepackage[colorlinks=true,linkcolor=blue,citecolor=blue, urlcolor=blue]{hyperref}
\usepackage[commentmarkup=uwave,defaultcolor=orange]{changes}
\usepackage{amsmath}
\usepackage{bm}

\usepackage{booktabs,multirow,makecell}

\begin{document}

\title{Exploration of Altermagnetism in $\mathrm{RuO_{2}}$}

\author{Yu-Xin Li}
\affiliation{Hubei Engineering Research Center of Weak Magnetic-field Detection, Department of Physics, China Three Gorges University, Yichang 443002, China}

\author{Yiyuan Chen}
\affiliation{Quantum Science Center of Guangdong-Hong Kong-Macao Greater Bay Area (Guangdong), Shenzhen 518045, China}

\author{Liqing Pan}
\affiliation{Hubei Engineering Research Center of Weak Magnetic-field Detection, Department of Physics, China Three Gorges University, Yichang 443002, China}

\author{Shuai Li}
\email{lis1277@ctgu.edu.cn}
\affiliation{Hubei Engineering Research Center of Weak Magnetic-field Detection, Department of Physics, China Three Gorges University, Yichang 443002, China}
\affiliation{State Key Laboratory of Quantum Functional Materials, Department of Physics,
 and Guangdong Basic Research Center of Excellence for Quantum Science,
 Southern University of Science and Technology (SUSTech), Shenzhen 518055, China}

\author{Song-Bo Zhang}
\email{songbozhang@ustc.edu.cn}
\affiliation{Hefei National Laboratory, Hefei, Anhui 230088, China}
\affiliation{School of Emerging Technology, University of Science and Technology of China, Hefei, Anhui 230026, China}

\author{Hai-Zhou Lu}
\email{luhz@sustech.edu.cn}
\affiliation{State Key Laboratory of Quantum Functional Materials, Department of Physics,
 and Guangdong Basic Research Center of Excellence for Quantum Science,
 Southern University of Science and Technology (SUSTech), Shenzhen 518055, China}
\affiliation{Quantum Science Center of Guangdong-Hong Kong-Macao Greater Bay Area (Guangdong), Shenzhen 518045, China}

\begin{abstract}
The fundamental role of magnetic materials in modern science and technology has driven a rapid surge in research on unconventional magnetism in recent years. Among these systems, altermagnets, which simultaneously exhibit zero net magnetization in real space and anisotropic spin splitting in momentum space, have garnered particular interest for both fundamental physics and technological applications. $\mathrm{RuO_{2}}$ stands as the pioneering and most extensively studied candidate in this class. While the intrinsic magnetic order of $\mathrm{RuO_{2}}$ remains a subject of active debate, numerous exotic phenomena characteristic of altermagnetism have been observed experimentally. In this review, we explore various facets of the altermagnetism through specific case studies in $\mathrm{RuO_{2}}$, including its crystal and magnetic structures, electronic band properties, and transport phenomena. We critically assess the debate surrounding the intrinsic magnetism in $\mathrm{RuO_{2}}$, incorporating evidence from altermagnetic signatures in transport with contrasting results from magnetic and spectroscopic measurements. Finally, we discuss possible future research directions on this topic.
\end{abstract}


\maketitle

\section{Introduction}

The ongoing quest for novel magnetic materials continues to drive fundamental research in condensed matter physics, offering avenues for next-generation spintronic technologies. Recent advances in this field have been led by the emergence of altermagnetism, a new classification of magnetic order distinct from traditional ferromagnetism and antiferromagnetism. Altermagnetism is characterized by a compensated (zero) net magnetization, similar to conventional antiferromagnets, yet exhibits spin-split electronic bands even without spin-orbit coupling, akin to those found in ferromagnets \cite{ref39,ref41,ref147}. This unique combination arises from specific crystal symmetries (rotation or mirror, rather than translation or inversion) that connect opposite-spin sublattices. These symmetries enforce sign-alternating, momentum-dependent spin polarization across the Brillouin zone, yielding spin-split bands without net magnetization. With the help of spin group symmetry analysis, altermagnetism has been explored in many materials, including $\mathrm{RuO_{2}}$ \cite{ref40,ref13,ref5,ref3,ref21}, MnTe \cite{ref159,ref155,ref157,ref160}, CrSb \cite{ref161,ref163,ref167}, $\mathrm{MnF_{2}}$ \cite{ref168,ref12,ref171}, $\mathrm{Mn_{5}Si_{3}}$ \cite{ref179,ref180,ref181}, $\mathrm{FeSb_{2}}$ \cite{ref183,ref182,ref184}, $\mathrm{CaMnO_{3}}$\cite{ref186,ref12}, $\mathrm{MnO_{2}}$ \cite{ref187}, $\mathrm{ReO_{2}}$ \cite{ref188}, $\mathrm{FeF_{2}}$ \cite{ref189}, $\mathrm{KV_{2}Se_{2}O}$ \cite{ref190,ref191}, $\mathrm{V_{2}Se_{2}O}$ \cite{ma2021multifunctional,zhang2025crystalsymmetrypaired} and $\kappa$-Cl \cite{ref193,ref192}. Furthermore, studies on altermagnets have led to discoveries of more unconventional magnetisms previously classified under antiferromagnetism \cite{ref176}, such as non-relativistic plaid-like spin splitting in a non-coplanar antiferromagnet \cite{ref177}, anomalous Hall antiferromagnets without spin splitting \cite{ref178}, and odd-parity spin splitting from coplanar antiferromagnets without spin-orbit coupling (SOC) \cite{zk69-k6b2}.
These new magnetic paradigms suggest novel routes for spin manipulation, holding significant promise for spintronic applications.

\begin{figure*}[t]
    \centering
    \includegraphics[width=1.0\linewidth]{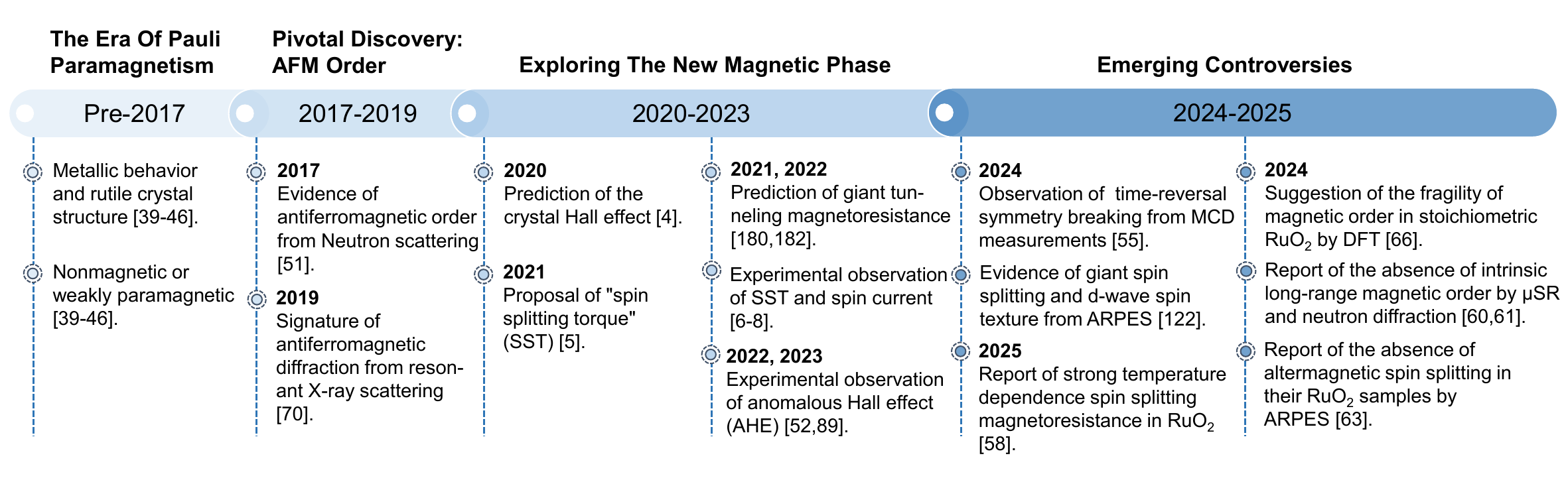}
    \caption{Timeline of research on the magnetism in $\mathrm{RuO_{2}}$. Key experimental and theoretical advances, highlighting reports of magnetic order and associated transport/spectroscopic signatures.
    }
    \label{fig:Timeline}
\end{figure*}

Ruthenium dioxide ($\mathrm{RuO_{2}}$) is among the earliest-proposed and most extensively studied altermagnet candidates. Historically, $\mathrm{RuO_{2}}$ was primarily categorized as a Pauli paramagnetic metal \cite{ref23,ref66,ref34,ref112,ref113,ref116,ref115,ref114}, and recognized for its utility in 
catalysis \cite{over2012surfacea,ref194} and energy storage \cite{ref195,ref196}. The surge of research interest in $\mathrm{RuO_{2}}$ in recent years is tightly tied to the renewed efforts to uncover its magnetic structure, as summarized in Fig.~\ref{fig:Timeline}. In 2017, a neutron scattering study reported antiferromagnetic ordering in $\mathrm{RuO_{2}}$ below $900$ K, stable down to at least $300$ K \cite{ref4}. This brings  $\mathrm{RuO_{2}}$ into the spotlight of unconventional magnetism research. In the same year when the concept of altermagnetism was introduced~\cite{ref41,ref147}, $\mathrm{RuO_{2}}$ was simultaneously predicted to be a prototypical altermagnetic candidate. This prediction was soon supported by the observation of both an anomalous Hall effect~\cite{ref40,ref11} and a spin-splitting torque effect~\cite{ref13,ref5,ref3,ref21}. These developments have spurred extensive investigations into $\mathrm{RuO_{2}}$, aiming to clarify the fundamental nature of altermagnetism and to harness the potential use of the zero-net-magnetization spin-splitting characteristics. 

While numerous reports continue to identify transport signatures in $\mathrm{RuO_{2}}$ that are consistent with altermagnetism~\cite{ref3,ref5,ref21,ref11,ref2,ref15,ref10,fan2024robust,he2025evidence,ref198,ref200}, more recent studies have introduced complexity by questioning the detection of the magnetic order in $\mathrm{RuO_{2}}$, thereby raising the debate on its true magnetic ground state. In particular, subsequent measurements using muon spin rotation and relaxation ($\mu$SR) \cite{ref17,ref22}, neutron diffraction \cite{ref8,ref22}, X-ray diffraction \cite{ref8} and angle-resolved photoemission spectroscopy (ARPES) \cite{ref27} have reported a non-magnetic state in $\mathrm{RuO_{2}}$. This consensus has further intensified research interest in this material. The influence of sample quality, epitaxial strain, stoichiometry deviations, and extrinsic disorder on the magnetic state of $\mathrm{RuO_{2}}$ is subsequently studied~\cite{ref46,ref36,ref42,ref17,ref22,ref19,ref27,ref71,ref8,ref49}. The ongoing efforts have advanced our understanding of the electronic and magnetic responses of $\mathrm{RuO_{2}}$.

This review presents a comprehensive and critical evaluation of the current understanding of the magnetic properties of $\mathrm{RuO_{2}}$. It is organized as follows. Section II overviews the crystal and magnetic structures of $\mathrm{RuO_{2}}$, focusing on the magnetic-structure characterizations that led to the initial controversies. Section III reviews the complex electronic structure of $\mathrm{RuO_{2}}$, including topological properties such as Dirac nodal lines and the altermagnetic spin splitting predicted by theory and observed in spectroscopic experiments. Sections IV surveys the charge, thermal and spin-related transport phenomena observed in $\mathrm{RuO_{2}}$, critically examining the transport features attributed to altermagnetism. In addition, Section V  surveys emergent phenomena such as strain-induced superconductivity and magneto-optical responses, emphasizing their connection to the underlying altermagmetic order. Finally, Section VI highlights unresolved aspects of magnetism in $\mathrm{RuO_{2}}$ and outlines promising directions for future research on altermagnetic materials.

\section{Crystal and magnetic structures}

\begin{figure*}[t]
    \centering
    \includegraphics[width=1.0\linewidth]{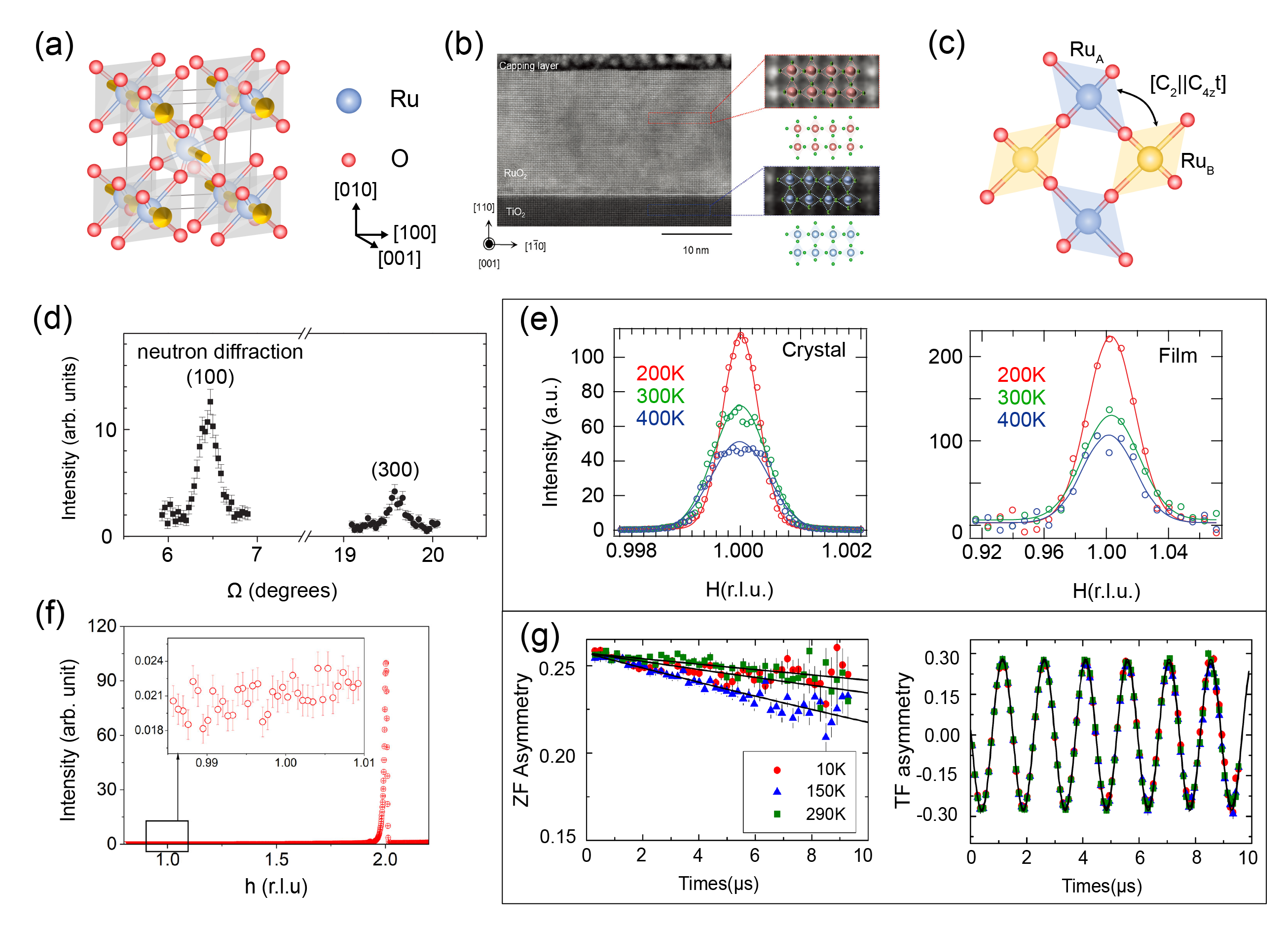}
    \caption{Structural and magnetic characterization of $\mathrm{RuO_{2}}$. (a) Crystal structure of $\mathrm{RuO_{2}}$. Ru atoms (blue), O atoms (red), and magnetic order along the $[001]$ direction. The yellow arrows mark the local magnetic moment at the Ru atoms. (b) Cross-sectional scanning transmission electron microscopy images of $\mathrm{RuO_{2}}$ films grown on $\mathrm{TiO_{2}}$ substrate, confirming the similarity of their structure \cite{ref71}. (c) Key spin-group symmetry element $[C_2||C_{4z}{\bm{t}}]$ of $\mathrm{RuO_{2}}$, which combines a two-fold spin-space rotation with a four-fold real-space rotation about the $z$-axis. $\mathrm{Ru_{A}}$ and $\mathrm{Ru_{B}}$ atoms are depicted in blue and yellow, respectively, and oxygen atoms are indicated in red. (d) Unpolarized neutron diffraction data taken at $295$ K. The results display enhanced intensity at odd-index reciprocal lattice points at room temperature, supporting the antiferromagnetic order and suggesting structural distortion in $\mathrm{RuO_{2}}$ \cite{ref4}. (e) Scattering intensities of resonant reflection (100) at different temperatures for (left) a bulk crystal and (right) a thin film $\mathrm{RuO_{2}}$. Both samples exhibit pronounced peaks near $H = 1$, persisting up to $400$ K, where $H$ denotes the reciprocal-space coordinate along the $(H,0,0)$ direction, and $H=1$ corresponding to a magnetic ordering wave vector $Q=(1,0,0)$, i.e., along the $[100]$ direction \cite{ref51}. (f) Neutron diffraction of $\mathrm{RuO_{2}}$ single crystal. In the magnified region of the figure, the $(100)$ reflection exhibits negligible intensity, with no detectable magnetic scattering signal observed \cite{ref22}. (g) Muon spin rotation and relaxation ($\mu$SR) in bulk $\mathrm{RuO_{2}}$. The curves were measured at different temperatures under zero (left) and weak (right) transverse magnetic fields \cite{ref22}.}
    \label{fig:Stucture-ND-XRD}
\end{figure*}
% NOA: d,e 

\subsection{Crystal structure}

$\mathrm{RuO_{2}}$ was regarded for a long time as a Pauli paramagnet \cite{ref23,ref66,ref34,ref112,ref113,ref116,ref115,ref114}. However, a pivotal study in 2017 \cite{ref4} revealed that it undergoes a structural distortion from rutile symmetry ($P4_2/mnm$, space group No.\,136) to two possible subgroups \cite{ref68} (space group No.\,77 and No.\,81) below $900$ K \cite{ref67,ref66,ref34}, accompanied by the onset of antiferromagnetic ordering that persists up to at least $300$ K. The rutile crystal structure of $\mathrm{RuO_{2}}$ is illustrated in Fig.~\ref{fig:Stucture-ND-XRD}(a). The magnetic Ru atoms are located at Wyckoff position $(0,0,0)$ and $(\frac{\text{1}}{2},\frac{\text{1}}{2},\frac{\text{1}}{2})$, highlighting the tetragonal symmetry and oxygen coordination that play a crucial role in the magnetic properties of $\mathrm{RuO_{2}}$.

The synthesis of high-purity samples with well-defined crystallographic features is essential for investigating the interplay between crystal structure and magnetism in $\mathrm{RuO_{2}}$, particularly given its subtle symmetry-breaking distortion and magnetic ordering. The synthesis method depends on the desired morphology. Single-crystalline $\mathrm{RuO_{2}}$ is typically synthesized by chemical vapor deposition (CVD) \cite{ref117,ref65,ref58,ref60,ref57,ref49}, yielding lattice constants of $a=b\approx4.49$ \AA~and $c\approx3.11$  \AA ~\cite{ref62,ref63,ref117,ref61,ref51,ref31,ref7}. In this method, $\mathrm{O_{2}}$ serves as the transport agent \cite{ref65} and the absence of additional reactants minimizes impurity incorporation, ensuring high-purity growth of bulk $\mathrm{RuO_{2}}$ crystals. Thin-film $\mathrm{RuO_{2}}$ can be synthesized using a broader range of techniques, including molecular beam epitaxy (MBE) \cite{ref23,ref69,ref25,ref21,ref22}, pulsed laser deposition (PLD) \cite{ref16,ref29,ref72,ref32,ref51,ref11,ref45,ref10,ref71}, physical vapor deposition (PVD) \cite{ref74}, reactive magnetron sputtering \cite{ref103,ref101,ref99,ref102,ref100,ref9}, atomic layer deposition (ALD) \cite{ref108,ref105,ref106,ref104,ref107}, and thermal laser epitaxy (TLE) \cite{ref109}. Among these, MBE, PLD, and PVD are particularly suitable for growing (110)-oriented $\mathrm{RuO_{2}}$ thin films on $\mathrm{TiO_{2}}$ substrates due to their lattice matching \cite{ref25,ref33,ref14}. Figure \ref{fig:Stucture-ND-XRD}(b) shows the structural coherence and high interface quality of $\mathrm{RuO_{2}}$ thin films grown on $\mathrm{TiO_{2}}$ substrate. PLD-grown $\mathrm{RuO_{2}}$ films have demonstrated a low resistivity at room temperature; however, they may contain approximately $10\%$ $\mathrm{RuO_{3}}$ \cite{ref16,ref29} impurities. Techniques such as reactive magnetron sputtering and ALD are highly sensitive to oxygen stoichiometry during $\mathrm{RuO_{2}}$ deposition \cite{ref99,ref105}, while TLE offers superior control over purity, enabling the growth of ultra-clean films \cite{ref109}. Compared with bulk crystals, $\mathrm{RuO_{2}}$ thin films typically exhibit slightly modified lattice constants, i.e., $a=b\approx4.59$ \AA~and $c\approx2.97$ \AA ~\cite{ref46,ref11,ref5,ref10,ref60}, likely due to epitaxial strain and interfacial effects.

\subsection{Evidence and analysis of magnetic Order}

Formerly considered as a Pauli paramagnet, $\mathrm{RuO_{2}}$ was revisited in a neutron scattering study in 2017 \cite{ref4}, which found a lattice distortion accompanied by a long-range, compensated magnetic order in the temperature range between 300 and 900 K. A subsequent resonant X-ray scattering (RXS) study in 2019 further confirmed this collinear long-range compensated magnetic order \cite{ref51}. These results were interpreted as evidence of a collinear antiferromagnetic order in $\mathrm{RuO_{2}}$. The unpolarized neutron diffraction suggested an ordered moment of approximately $0.23$ $\mu_B$ per Ru atom, whereas refined polarized neutron scattering revealed a substantially reduced moment of about $0.05$ $\mu_B$~\cite{ref4}. Figure~\ref{fig:Stucture-ND-XRD}(d) displays enhanced intensity at odd-index reciprocal lattice points at room temperature, consistent with long-range compensated magnetic order and a concomitant structural distortion \cite{ref4}. DFT calculations incorporating the previously overlooked effects of on-site Coulomb interactions among the Ru 4d orbitals successfully reproduced the magnetic order found in the experiments. Calculations also showed that the compensated magnetic order state always has lower energy than the ferromagnetic state~\cite{ref4}. RXS measurements (Fig.~\ref{fig:Stucture-ND-XRD}(e)) further revealed the diffraction peak intensities in both $\mathrm{RuO_{2}}$ bulk crystals and thin films over a wide range of temperatures \cite{ref51}. Both samples exhibit pronounced peaks near $H=1$, where $H$ denotes the reciprocal-space coordinate along the $(H,0,0)$ direction, and $H=1$ corresponds to a magnetic order wave vector $Q=(1,0,0)$, i.e., along the $[100]$ direction. The persistence of these peaks up to $400$ K confirms the room-temperature long-range compensated magnetism in $\mathrm{RuO_{2}}$.

Guided by experimental evidence of antiferromagnetism, DFT calculations further clarify the magnetic structure and its microscopic origin \cite{ref4, ref52}. In particular, they revealed that the magnetic moment of Ru atom arises from a Fermi surface instability, pointing to an itinerant rather than localized-moment mechanism. Recent dynamic interfacial measurements on $\mathrm{RuO_{2}}$-based heterostructures further demonstrate that interfacial exchange coupling can convert a commensurate SDW into an incommensurate SDW, firmly supporting the itinerant SDW mechanism of magnetism in $\mathrm{RuO_{2}}$ \cite{feng2024incommensurate}. Such incommensurate SDW may promote the formation of an orbital Kondo effect \cite{yeh2020oxygen}. Notably, the antiferromagnetic order in $\mathrm{RuO_{2}}$ breaks time-reversal symmetry, which suppresses the oxygen-vacancy–induced orbital Kondo effect from a potential two-channel regime to a one-channel regime. This interplay between itinerant magnetism, symmetry breaking, and Kondo physics highlights the rich landscape of correlated behavior in $\mathrm{RuO_{2}}$. 

The total magnetic moment can be expressed as $\mathbf{m}=\mathbf{m}_a+\mathbf{m}_b$, where $\mathbf{m}_a$ and $\mathbf{m}_b$ represent the magnetic moments at Ru$_A$ and Ru$_B$ atoms, respectively. In the absence of SOC, the magnetic moments perfectly compensate, consistent with the Ru-projected density of states \cite{ref40}. The Néel vector, which characterizes the direction of antiferromagnetic order, is oriented along the crystallographic axis $[001]$ or $[00\bar1]$ \cite{ref5,ref12,ref51}. This prediction has been directly confirmed by X-ray magnetic linear dichroism (XMLD) measurements. Explicitly, in $\mathrm{RuO_{2}}$ thin films the Néel vector is nearly parallel to the $[001]$ axis \cite{ref9}, with a slight deviation \cite{ref21}. This observation aligns with the theoretical framework of altermagnetism, which predicts both compensated magnetic moments and spin-split bands \cite{ref13}. 

The magnetic and symmetry properties of $\mathrm{RuO_{2}}$ underpin its classification as an altermagnet and clarify its microscopic origins of its spin splitting in the electronic band structure. Unlike ferromagnets and noncollinear antiferromagnets, which break the combined symmetry of time-reversal and spatial inversion (or translation) through their internal spin configurations \cite{ref6,ref24}, $\mathrm{RuO_{2}}$ does so via an anisotropic magnetization density within each spin sublattice, induced by the crystal arrangement of nonmagnetic oxygen atoms \cite{ref40}. This symmetry breaking results in anisotropic spin-split bands despite zero net magnetic magnetization~\cite{ref13}. The magnetic anisotropy of $\mathrm{RuO_{2}}$ has been systematically studied: the magnetic symmetry depends on the Néel vector orientation. For the Néel vector $\bf {n}$ along the $[100]$ axis, the symmetry group is $Pnn'm'$, whereas for $\bf {n}$ along $[110]$, it is $Cnn'm'$. The specific magnetic symmetry has significant physical consequences, such as notably the anisotropic anomalous Hall conductivity \cite{ref40}.

In parallel, the magnetic properties of $\mathrm{RuO_{2}}$ have been analyzed from a symmetry-based perspective. To describe this novel magnetic phase properly, spin space groups have been employed, which extend crystallographic symmetries to capture the magnetic structure of collinear antiferromagnets \cite{ref39,ref41,ref12}. $\mathrm{RuO_{2}}$, the prototypical altermagnet candidate, is described by the spin group $\mathbf{R}_{s}^{III}=[E||\mathbf{H}]+[C_2||\mathbf{G-H}]$, where $E$ denotes the spin-space identity and $\mathbf{H}$ is a subgroup containing half of the real-space operations (including the identity) of the nonmagnetic crystallographic group $\mathbf{G}$. The remaining operations, $\mathbf{G-H}$, can be expressed as $A\mathbf{H}$ , where $A$ is a (proper or improper) real-space rotation \cite{ref41}. This symmetry group implies that opposite-spin sublattices are related by rotational, not by translational or inversion operations. Consequently, even in the nonrelativistic limit, the band structure exhibits spin splitting \cite{ref39}. 

A crucial spin group symmetry of $\mathrm{RuO_{2}}$ is $[C_{2}||C_{4z}\mathbf{t}]$, which combines a two-fold spin-space rotation with a four-fold real-space rotation about the $z$ axis. By contrast, in relativistic magnetic groups the allowed operations act simultaneously in spin and real spaces because spin is tied to lattice rotations via SOC. As illustrated in Fig. \ref{fig:Stucture-ND-XRD}(c), where $C_{2}$ represents a $180^{\circ}$ spin-space rotation about an axis perpendicular to the spins. On the right side of the double vertical bar, $C_{4z}\mathbf{t}$ denotes a four-fold real-space rotation combined with a lattice translation. 

%\begin{figure*}[t]
   % \centering
   % \includegraphics[width=1.0\linewidth]{section2 graphics/Non-AM.png}
   % \caption{(a) Hole doping induced by Ru vacancies. The color gradient represents the transition from nonmagnetic (dark) to magnetic (light) states. Undoped $\mathrm{RuO_{2}}$ remains nonmagnetic, and the magnetic order emerges under Ru vacancies \cite{ref42}. (b) Neutron diffraction of $\mathrm{RuO_{2}}$ single crystal. In the magnified region of the figure, the $(100)$ reflection exhibits negligible intensity, with no detectable magnetic scattering signal observed \cite{ref22}. (c) Muon spin rotation and relaxation in bulk $\mathrm{RuO_{2}}$. The curves were measured at different temperatures under zero (top) and weak (bottom) transverse magnetic fields \cite{ref22}. (d) XRD powder data for a $\mathrm{RuO_{2}}$ sample. No impurity peaks were found, the sample was an undistorted rutile structure described by these data \cite{ref8}.}
   % \label{fig:Non-AM}
%\end{figure*}
% NOA: a 

\subsection{Reports of non-magnetic ground states}

Despite exciting advances in theory and experiment, earlier studies on the magnetic properties of $\mathrm{RuO_{2}}$, which suggested an absence of intrinsic magnetism, underscore the need to reevaluate its magnetic character \cite{ref23,ref66,ref34,ref112,ref113,ref116,ref115,ref114}. More recent investigations have further raised concerns about sample quality and measurement sensitivity \cite{ref42,ref17,ref22,ref19,ref27,ref8,ref49}. A systematic DFT study reported that stoichiometric $\mathrm{RuO_{2}}$ exhibits a negligible intrinsic magnetic moment, and the magnetism can emerge through hole doping induced by Ru vacancies \cite{ref42}. Generally, DFT+U is used for calculating the electronic properties of $\mathrm{RuO_{2}}$, where U accounts for the strong correlation effect arising from Coulomb interaction in Ru 4d orbitals. The magnetism emerges in $\mathrm{RuO_{2}}$ when the parameter U greater than 1 eV is used, which is too large for the realistically experienced value. However, it is found that hole doping can effectively reduce the value of U needed for magnetism \cite{ref42}. Later, as the origin of magnetism in $\mathrm{RuO_{2}}$ was further explored, it was found that $\mathrm{RuO_{2}}$ is in proximity to a quantum phase transition. Parameters like hole doping and epitaxial strain can modulate the quasiparticle interactions near the Fermi surface, leading to magnetism in $\mathrm{RuO_{2}}$ \cite{WuCJ2025}.

The intrinsic magnetic state of $\mathrm{RuO_{2}}$ has been further examined using highly sensitive experimental techniques such as muon spin rotation and relaxation ($\mu$SR) which directly probe local magnetic environment \cite{ref17,ref22}. In Ref.~\cite{ref17}, a $100\%$ spin-polarized $\mu^+$ beam was injected into the sample, and the $\mu^+$ spin was monitored by the time-dependent asymmetry $A(t)$ of the positron. The $\mu$SR time spectra were measured under zero-field conditions between $5$ and $400$ K and show a slow exponential decay without a sinusoidal component of Larmor precession signal, indicating the absence of a uniform internal magnetic field $B_{\text{loc}}$. Moreover, a weak longitudinal magnetic field of $1$ mT nearly suppresses this relaxation, confirming the quasistatic nature of $B_{\text{loc}}$ and excluding the presence of strong dynamic fields. Taken together, these findings indicate that $\mathrm{RuO_{2}}$ behaves as a nonmagnetic metal under stoichiometric conditions.

Combined single crystal neutron diffraction and $\mu$SR measurements provide strong evidence against intrinsic long-range magnetic order in $\mathrm{RuO_{2}}$ \cite{ref22}. In Fig. \ref{fig:Stucture-ND-XRD}(f), the horizontal axis $h$ represents the coordinate in reciprocal space, specifically corresponding to the reciprocal lattice along the $(h, 0, 0)$ direction. As shown in the magnified region, the intensity of the $(1,0,0)$ reflection is markedly weak, in sharp contrast to earlier reports of strong magnetic scattering signals. The earlier observation of the $(1,0,0)$ peak was later shown to depend strongly on the scattering angle, a behavior inconsistent with that of a true Bragg reflection, and was attributed to multiple scattering, mainly double scattering, in accordance with the Renninger effect rather than intrinsic magnetic order. The $\mu$SR results corroborate this conclusion. As illustrated in Fig. \ref{fig:Stucture-ND-XRD}(g), at zero field, no measurable relaxation or depolarization was observed at any temperature, while in a weak transverse field ($5$ mT), the Larmor precession amplitude is small and shows no significant temperature dependence. These observations rule out any sizable internal magnetic field. Quantitative analysis sets an upper limit of $\sim 10^{-4}$ $\mu_B$  per Ru atom on the magnetic moment, effectively excluding long-range magnetic order. Beyond bulk-phase measurements, low-energy $\mu$SR experiments were conducted to probe the magnetism of $\mathrm{RuO_{2}}$ thin films \cite{ref22,akashdeep2025surfacelocalized}. The distinction between signals originating from the film and those from the substate was achived through energy-dependent intensity analysis. Both studies consistently report the absence of long-range magnetic order in the thin films. These results suggest that the previously reported altermagnetic features may instead originate from extrinsic effects (e.g., multiple scattering or structural defects). They highlight the importance of sample quality in determining the magnetic ground state of $\mathrm{RuO_{2}}$ and the need for precise control over synthesis conditions.

Bulk-sensitive measurements paint a similarly nonmagnetic picture. Optical conductivity data obtained from Kramers-Kr\"onig transformations of reflectance spectra, closely match nonmagnetic band structure calculations, suggesting that the bulk electronic structure lacks clear signatures of altermagnetic order \cite{ref49}. Both experimental and computed optical conductivities display a sharp increase near a photon energy of $0.1$ eV and a pronounced minimum around $2$ eV, reinforcing the nonmagnetic model for bulk $\mathrm{RuO_{2}}$. This suggests that previously reported altermagnetic signatures may arise from strain or non-stoichiometry effects in thin films rather than intrinsic bulk properties \cite{ref36,ref70}.

To illustrate the impact of synthesis quality on magnetic properties, a recent study \cite{ref8} focused on high-purity single-crystalline $\mathrm{RuO_{2}}$ synthesized via chemical vapor transport, aiming to eliminate extrinsic defects (e.g., vacancies) and re-examine the intrinsic ground state. The samples were subsequently characterized by powder X-ray diffraction (XRD), resistivity and magnetic measurements, single-crystal XRD, and neutron diffraction. Resistivity and magnetic susceptibility measurements indicated metallic behaviors consistent with near-perfect stoichiometry, while magnetic measurements showed no evidence of magnetic phase transitions. Powder XRD confirmed an undistorted rutile structure without detectable impurity phases. Both X-ray and neutron diffraction analyses demonstrated that $\mathrm{RuO_{2}}$ retains an ideal rutile structure at low temperatures, with negligible Ru vacancies. Notably, polarized neutron diffraction analysis ruled out the previously proposed antiferromagnetic structure with magnetic moments exceeding $0.01$ $\mu_B$, suggesting an absence of antiferromagnetic order in stoichiometric $\mathrm{RuO_{2}}$. Complementary measurements on ultra–high-purity $\mathrm{RuO_{2}}$ single crystals, likewise detected no evidence of magnetic order down to $1.8$ K \cite{paul2025growth}. In parallel, magnetic Raman spectroscopy measurements \cite{abel2025probing} demonstrating that a magnon mode in $\mathrm{RuO_{2}}$ films is only observable the material is interfaced with a ferromagnetic $\mathrm{NiFe}$ layer, suggesting that any reported magnetic signatures in $\mathrm{RuO_{2}}$ likely originate from interface effects rather than a bulk property. This interpretation is consistent with first-principles calculations demonstrating that the $\mathrm{RuO_{2}}$$(110)$ surface can spontaneously develop substantial magnetization \cite{ho2025symmetrybreaking}. The origin of this surface magnetism is attributed to symmetry reduction at the surface, which reconstructs the local electronic structure and stabilizes a magnetic state absent in the bulk.

Recent studies on ultrathin $\mathrm{RuO_{2}}$ have revealed a range of intriguing emergent phenomena. Ab initio calculations \cite{brahimi2024confinementinduced} suggest that strain relaxation in $\mathrm{RuO_{2}}$ thin films substantially modifies the electronic structure. This effect, analogous to that of a Hubbard-U correction, is predicted to induce altermagnetism without requiring an explicit U correction. Complementary experimental studies on strained ultrathin ($\sim 2.7$ nm) $\mathrm{RuO_{2}}$ films identified a complex spin texture. Comprising both mirror-odd and mirror-even parity components, with the latter providing direct evidence for time-reversal symmetry breaking \cite{zhang2025observation}. In contrast, spin-polarized scanning tunneling microscope measurements on ultrathin $\mathrm{RuO_{2}}$ $(110)$ films reported no indication of surface magnetic instability \cite{kessler2025moireassisted}, suggesting that the emergence of magnetism in reduced dimensions is highly sensitive to epitaxial strain and structural relaxation.

In summary, while the discovery of symmetry-allowed altermagnetism in $\mathrm{RuO_{2}}$ has motivated extensive theoretical and experimental efforts, the existence of intrinsic long-range magnetic order remains under active debate.

\section{Electronic structure}

\subsection{Topological properties}

\begin{figure*}[t]
    \centering
    \includegraphics[width=1.0\linewidth]{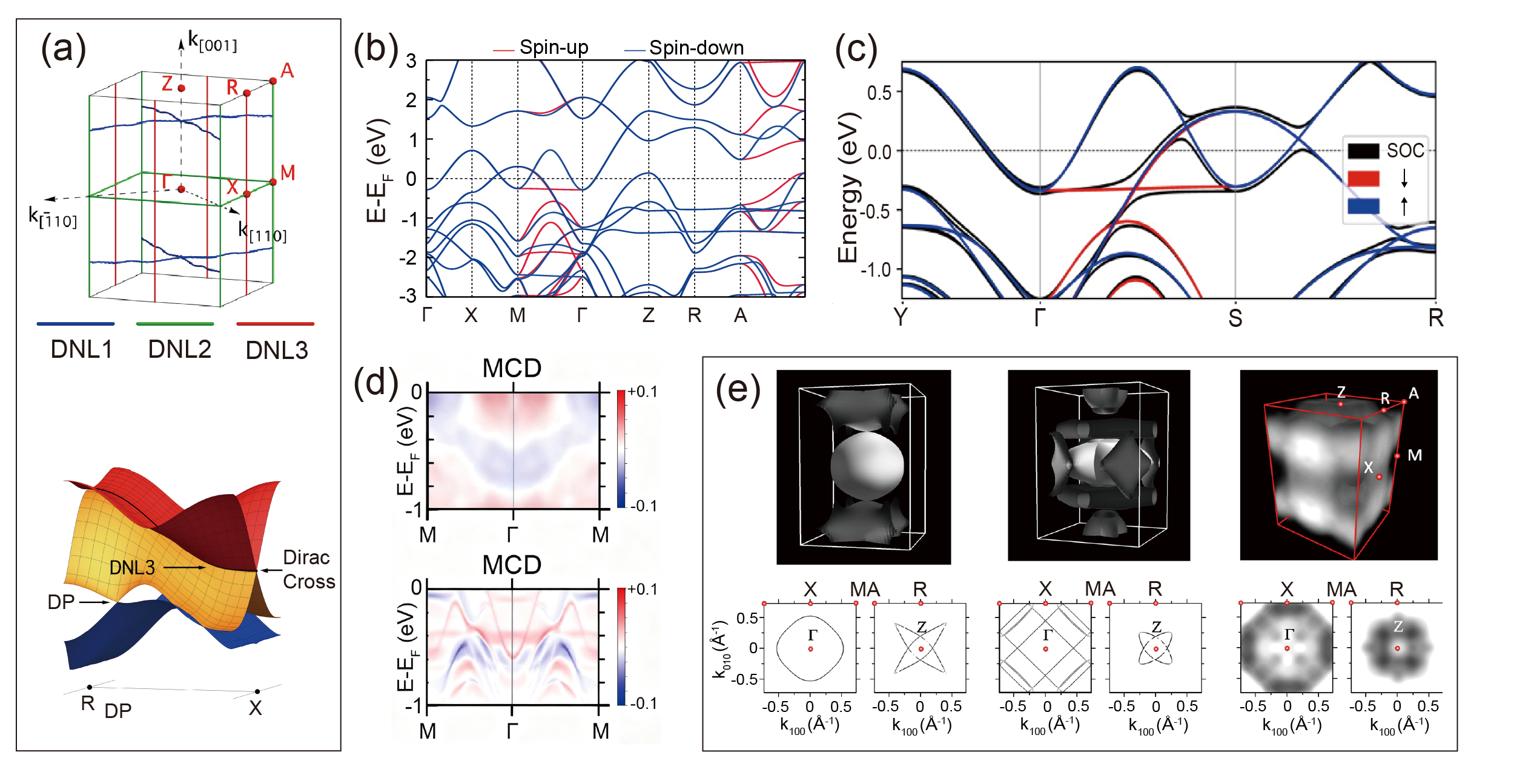}
    \caption{Topological band structure and spin-splitting features in $\mathrm{RuO_{2}}$. (a) Dirac nodal lines (DNLs) in $\mathrm{RuO_{2}}$ \cite{ref53}. The calculated k-space trajectories of the three DNLs in the $\mathrm{RuO_{2}}$ Brillouin region (top). DNL3 shown in the DFT band structure model of $\mathrm{RuO_{2}}$ crosses the fourfold band along XR and a Dirac point (bottom). (b) Calculated band structure of $\mathrm{RuO_{2}}$, showing the crossing nodes and spin splitting \cite{ref20}. (c) Energy bands of $\mathrm{RuO_{2}}$ near the Fermi level, without (red and blue) and with (black) relativistic SOC \cite{ref41}. (d) Magnetic circular dichroism (top) and DFT calculations (bottom) of $\mathrm{RuO_{2}}$ spin polarization \cite{ref10}. (e) 3D Fermi surfaces calculated for the paramagnetic and altermagnetic phases are compared with the measured Fermi surface \cite{ref10}. The left parts are the calculated Fermi surface in 3D (top) and projected to 2D planes (bottom) in the paramagnetic phase of $\mathrm{RuO_{2}}$, the middle parts corresponds to the calculation of altermagnetic phase. The right parts are the Fermi surface obtained from topographic mapping experiments of $\mathrm{RuO_{2}}$ through the 3D Brillouin zone (top), and the photoelectron intensity at the plane Fermi energy (bottom).}
    \label{fig:Spinsplit-DNLs-MCD}
\end{figure*}
% NOA: a

The electronic structure of $\mathrm{RuO_{2}}$, especially the band topology near the Fermi energy, plays an important role in determining its magnetic properties. DFT calculations revealed a complex band structure dominated by the Ru $4d$ ($d_{xy}$, $d_{xz}$, and $d_{yz}$) orbitals~\cite{ref39,ref36,ref111}. These orbitals generate multiple band crossings near the Fermi level and spin splitting characteristic of altermagnetism. The interaction between these orbitals and the underlying crystal symmetry plays a crucial role in shaping the momentum-dependent spin polarization observed in the altermagnetic state \cite{ref52,ref14,ref39,ref36,ref111}.

\begin{figure*}[t]
    \centering
    \includegraphics[width=1.0\linewidth]{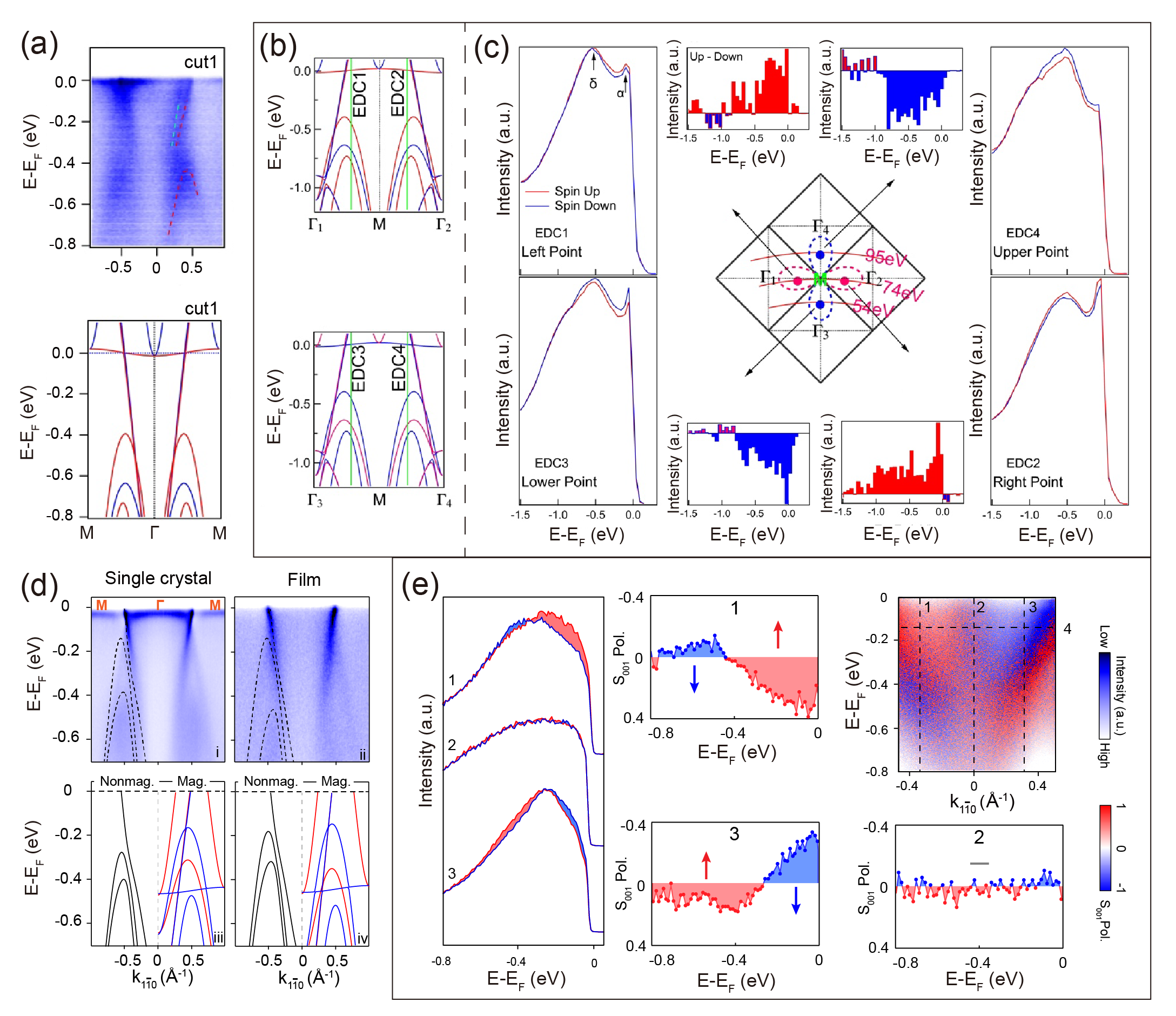} 
    \caption{Comparison of ARPES and spin-resolved ARPES signatures in $\mathrm{RuO_{2}}$. (a) The existence of spin-polarized bands in $\mathrm{RuO_{2}}$. ARPES spectra along $M$-$\Gamma$-$M$ momentum cut (top) and corresponding calculated spin-resolved bands (bottom)~\cite{ref28}. (b) Calculated spin-polarized band structures along two orthogonal directions, $\Gamma_{1}$-$M$-$\Gamma_{2}$ and $\Gamma_{3}$-$M$-$\Gamma_{4}$~\cite{ref28}. (c)Schematic of the Brillouin-zone cross section (center) indicating the four measurement points used for extracting the spin-resolved energy distribution curves (EDCs). The surrounding panels show the spin-polarized EDCs at the left (EDC1), right (EDC2), lower (EDC3), and upper (EDC4) points, respectively. The differences between spin-up and spin-down intensities at these points are plotted in the adjacent upper and lower histograms~\cite{ref28}. (d) Comparison of ARPES spectra (top) and calculation (bottom) for the energy band structure of single crystal and thin film $\mathrm{RuO_{2}}$~\cite{ref27}. Panels i and iii correspond to the single crystal, while panels ii and iv correspond to the thin film.  Neither sample exhibited the spin splitting predicted for the altermagnetic phase. (e) Spin-resolved EDCs measured at three selected momenta (1–3) and their corresponding spin polarization ($S_{y}$) spectra. The rightmost panel presents the spin-resolved ARPES intensity map, showing the momentum–energy distribution of the spin polarization~\cite{ref27}. }
    \label{fig:ARPES-AM_NAM}
\end{figure*}
% NOA: c

Symmetry and topology-based analyses predicted multiple Dirac nodal lines (DNLs) in $\mathrm{RuO_{2}}$ \cite{ref84,ref81,ref82,ref85,ref78,ref79,ref86,ref83,ref80,ref87}, which are continuous, symmetry-protected lines of band crossings. %However, the inclusion of SOC modifies this nodal topology by introducing band gaps or changing the protection mechanisms.
In particular, DNL1 lies in the $(110)$ and $(\bar110)$ mirror planes and is stabilized by time-reversal, space inversion, and mirror symmetries \cite{ref88}. DNL2 runs along the Brillouin zone boundaries at $k_x=\pi/a$ and $k_y=\pi/a$, protected by a nonsymmorphic (sliding) mirror symmetry \cite{ref89,ref88,ref87}. Additionally, DFT calculations on the $(110)$ surface predicted an additional DNL3 along the XR direction (Fig.~\ref{fig:Spinsplit-DNLs-MCD}(a)), which was later confirmed by ARPES measurements and found to be robust against strong SOC, giving rise to flat-band surface states that connect adjacent nodal lines \cite{ref53,ref210}. It is particularly significant, with implications for surface reactivity, spin Hall conductivity, and macroscopic magnetic response. Figure \ref{fig:Spinsplit-DNLs-MCD}(b) and (c) present the band structure of $\mathrm{RuO_{2}}$ found from DFT calculations \cite{ref20,ref41}.  Topological classification has revolutionized the understanding of matter, and altermagnetism has unveiled a new magnetic order. The study of the interplay between altermagnetism and topological states is an emerging frontier. An important feature of their combination is the coexistence of Dirac/Weyl node crossing within the same spin-polarized energy band \cite{ref76}. For two-dimensional (2D) systems, it is possible to have quantized and spin-polarized transport. For three-dimensional (3D) systems, such as $\mathrm{RuO_{2}}$, it is proposed that spin-polarized Fermi arcs can exist. These spin-polarized topological surface/edge states are promising targets for further exploration.

Spin-orbit coupling modifies the nodal topology. In many altermagnetic models, without SOC, band crossings between states of the same spin but on different sublattices yield DNLs in two dimensions (2D) and Weyl nodal lines in three dimensions (3D) \cite{ref92,ref95,ref94,ref90,ref91,ref93}. With SOC, the DNL1 of the 2D altermagnets gaps and evolves into Weyl nodal lines. For $\mathrm{RuO_{2}}$ with rutile structure, these Weyl nodal lines remain symmetry-protected and persist in the $k_z=\pi$ plane \cite{ref76}. Furthermore, the origin of the magnetic order of $\mathrm{RuO_{2}}$, which lies in the Fermi surface instability, is linked to the presence of a nodal line close to the Fermi level \cite{ref52}.

The band structure of $\mathrm{RuO_{2}}$ near the Fermi energy can be described by an effective Hamiltonian \cite{ref111},
\begin{equation}
H=\varepsilon_{0, \mathbf{k}}+t_{x, \mathbf{k}} \tau_x+t_{z, \mathbf{k}} \tau_z+\tau_y \bm{\lambda}_{\mathbf{k}} \cdot \bm{\sigma}+\tau_z \bm{J} \cdot \bm{\sigma},
\end{equation}
where ${\bf{k}}=(k_x,k_y,k_z)$ is the 3D crystal momentum, $\bm{\sigma}=(\sigma_x,\sigma_y,\sigma_z)$ and $\bm{\tau}=(\tau_{x},\tau_y,\tau_z)$ are Pauli matrices acting in spin and sublattice spaces, respectively.  %$\varepsilon_{0, \mathbf{k}}$ denotes the dispersion independent of the sublattice, $t_{x, \mathbf{k}}$ and $t_{z, \mathbf{k}}$ represent inter- and intra-sublattice hopping, respectively, $\bm{\lambda}_{\mathbf{k}}$ accounts for SOC, and $\bm{J}$ represents the altermagnetic exchange field. 
$\varepsilon_{0, \mathbf{k}}$ denotes the dispersion independent of sublattice and spin, and it reads
%\begin{equation}
%\begin{aligned}
%\varepsilon_{0, \mathbf{k}}= & t_1\left(\cos k_x+\cos k_y\right)+t_2 \cos k_z+t_3 \cos k_x \cos k_y \\
%& +t_4\left(\cos k_x+\cos k_y\right) \cos k_z \notag \\
%& +t_5 \cos k_x \cos k_y \cos k_z -\mu ,
%\end{aligned}
%\end{equation}
\begin{equation}
\begin{aligned}
\varepsilon_{0, \mathbf{k}}= & (t_1 +t_4 \cos k_z )\left(\cos k_x+\cos k_y\right)+t_2 \cos k_z \\
&  -\mu + (t_3 +t_5 \cos k_z) \cos k_x \cos k_y .
\end{aligned}
\end{equation}
%the hopping coefficients
The inter- and intra-sublattice hoppings are
\begin{equation}
\begin{aligned}
& t_{x, \mathbf{k}}=t_8 \cos \frac{k_x}{2} \cos \frac{k_y}{2} \cos \frac{k_z}{2}, \\
& t_{z, \mathbf{k}}=(t_6+t_7 \cos k_z) \sin k_x \sin k_y,
\end{aligned}
\end{equation}
and the SOC term $\bm{\lambda}_{\mathbf{k}}=(\lambda_x,\lambda_y,\lambda_z)$ is 
\begin{equation}
\begin{aligned}
& \lambda_{x, \mathbf{k}}=\lambda \sin \frac{k_z}{2} \sin \frac{k_x}{2} \cos \frac{k_y}{2}, \\
& \lambda_{y, \mathbf{k}}=-\lambda \sin \frac{k_z}{2} \sin \frac{k_y}{2} \cos \frac{k_x}{2}, \\
& \lambda_{z, \mathbf{k}}=\lambda_z \cos \frac{k_z}{2} \cos \frac{k_x}{2} \cos \frac{k_y}{2}\left(\cos k_x-\cos k_y\right).
\end{aligned}
\end{equation}
%The energy bands are given by
%\begin{equation}
%\begin{split}
%E_{\alpha=\pm,\beta=\pm} & = %\varepsilon_{0,\mathbf{k}} + \alpha \Big[  %t_{x,\mathbf{k}}^{2} + t_{z,\mathbf{k}}^{2} + %|\bm{\lambda}_{\mathbf{k}}|^{2} + |\bm{J}|^{2} \\
%&\qquad + 2 \beta \sqrt{t_{z, \mathbf{k}}^2 %|\bm{J}|^2 + |\bm{\lambda}_\mathbf{k} \times %\bm{J}|^2} \Big]^{1/2}.
%\end{split}
%\end{equation}
%In the absence of SOC ($\bm{\lambda}_{\mathbf{k}}=0$), degeneracies arise on a Weyl plane defined by $t_{z, \mathbf{k}}=0$ and a Weyl line defined by $t_{z, \mathbf{k}}=0$, and $t_{z, \mathbf{k}}=\pm|\bm{J}|$. These degeneracies significantly influence the low-energy band structure, dictating the topologically protected crossings and affecting the electronic transport properties. When SOC is present, the Weyl plane is mostly gaped, leaving either a Weyl line or Weyl point, depending on the specific band interactions.
It is worth noting that this Hamiltonian is the symmetry-allowed minimal effective low-energy model for $\mathrm{RuO_{2}}$. It describes the energy bands of $\mathrm{RuO_{2}}$ near the Fermi energy, where the altermagnetic spin splitting is controlled by a single parameter. It serves as an important basis for advanced model studies of $\mathrm{RuO_{2}}$, which are currently limited.

%Figure \ref{fig:Spinsplit-DNLs-MCD}(b) and (c) present the band structure of $\mathrm{RuO_{2}}$ found from DFT calculations \cite{ref20,ref41}. 

The interplay between SOC and nodal topology not only affects band crossings but also leads to finite Berry curvature, which in turn governs transport in $\mathrm{RuO_{2}}$. Although $\mathrm{RuO_{2}}$ consists of nominally nonmagnetic elements and exhibits a collinear antiferromagnetic ground state, the asymmetric arrangement of oxygen atoms breaks time-reversal symmetry in the electronic band structure, enabling finite Berry curvature \cite{ref40,ref50,ref37}. The Berry curvature, which determines the geometric phase of Bloch electrons \cite{ref98,ref96,ref97,ref40,ref39,ref41,ref54}, can be calculated by
\begin{equation}
{\mathbf{\Omega}}_n(\mathbf{k})=\operatorname{Im} \sum_{n^{\prime} \neq n} \frac{\langle n \mathbf{k}| \nabla_\mathbf{k} H(\mathbf{k})\left|n^{\prime} \mathbf{k}\right\rangle \times\left\langle n^{\prime} \mathbf{k}\right| \nabla_\mathbf{k} H(\mathbf{k})|n \mathbf{k}\rangle}{\left(E_{n \mathbf{k}}-E_{n^{\prime} \mathbf{k}}\right)^2}.
\end{equation}
where $\lvert n\mathbf{k}\rangle$ are cell-periodic Bloch eigenstates of $H(\mathbf{k})$ with eigenvalues $E_{n\mathbf{k}}$. 
In RuO$_2$, the distortions in the magnetization density induced by oxygen asymmetry result in finite Berry curvature, enabling macroscopic phenomena such as the anomalous Hall effect~\cite{ref73,ref13}. These effects are most pronounced near band crossings, such as Dirac or Weyl nodes.

\subsection{Altermagnetic band splitting}

Theoretical studies indicated that $\mathrm{RuO_{2}}$ hosts spin-split bands analogous to those in ferromagnets; however, the spin polarization alternates in momentum space--a defining signature of altermagnetism that preserves zero net magnetization. DFT calculations consistently predicted this momentum-dependent spin splitting in $\mathrm{RuO_{2}}$ \cite{ref55,ref56,ref52,ref59,ref54}. A recent study calculated the alternating spin splitting up to $1.54$ eV \cite{ref28}, in line with earlier predictions \cite{ref40,ref39,ref41,ref54,ref31}. This magnitude is significantly larger than those observed in most other altermagnets \cite{ref64} and comparable to spin splitting in typical ferromagnets. 

Recent magnetic circular dichroism (MCD) and topographic mapping measurements provided more direct evidence of momentum-dependent spin polarization in $\mathrm{RuO_{2}}$, supporting the altermagnetic scenario. As shown in Fig.~\ref{fig:Spinsplit-DNLs-MCD}(d), both theoretical calculations and experimental MCD signals indicate a strong time-reversal symmetry breaking \cite{ref10}. To better verify the altermagnetic properties of $\mathrm{RuO_{2}}$, the key theoretically predicted features in paramagnetic and altermagnetic phases were compared with the experimental results~\cite{ref10}. In the paramagnetic state (Fig.~\ref{fig:Spinsplit-DNLs-MCD}(e), left), the Fermi surface exhibits the time-reversal symmetry, reflecting the absence of any intrinsic spin polarization. In contrast, in the altermagnetic phase (Fig.~\ref{fig:Spinsplit-DNLs-MCD}(e), middle), a clear band splitting and pronounced momentum-space asymmetry emerge. Notably, a Brillouin zone crossing develops along $\mathit{\Gamma}$-$M$ that is absent in the paramagnetic phase. The DFT-predicted Fermi-surface spin polarization closely matches the 3D Brillouin zone topographic mapping obtained experimentally (Fig.~\ref{fig:Spinsplit-DNLs-MCD}(e), right). This further confirms the alternating spin polarization in $\mathrm{RuO_{2}}$. 

\begin{figure*}
    \centering
    \includegraphics[width=1.0\linewidth]{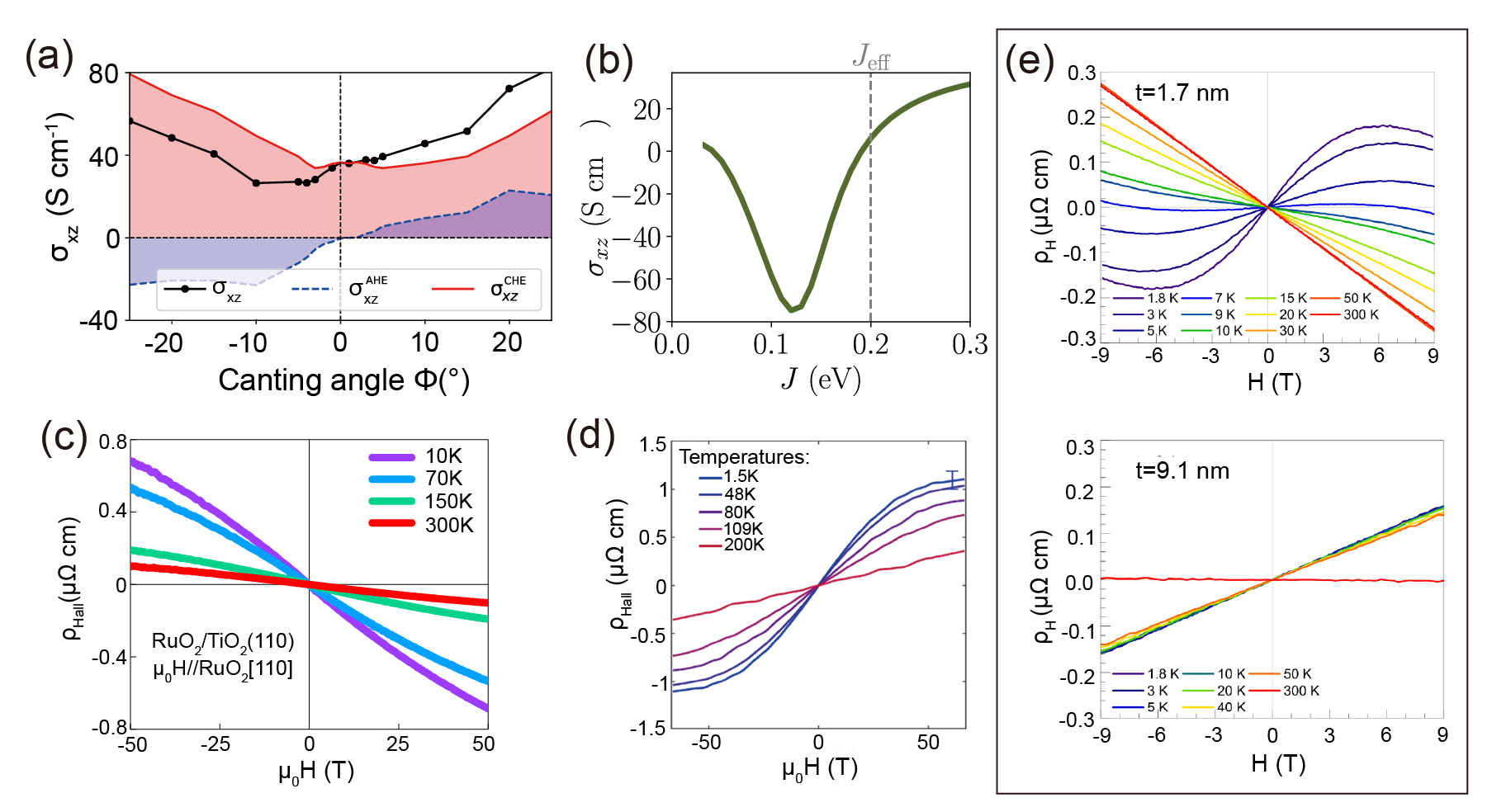}
    \caption{Symmetry-tuned Hall responses in $\mathrm{RuO_{2}}$. (a) Calculated Hall conductivity ($\sigma_{xz}$) and the Hall conductivity components ($\sigma_{xz}^{\mathrm{AHE}},\sigma_{xz}^{\mathrm{CHE}}$) as functions of the sublattice magnetization (canting angle) \cite{ref40}.(b) Calculated sign and magnitude dependence of $\sigma_{xz}$ on the altermagnetic strength $J$ \cite{ref111}. (c) Temperature-dependent Hall resistivity of $(110)$ film \cite{ref11}. (d)  Saturation of the Hall resistivity \cite{ref45}. (e) Temperature-dependent Hall resistivity of ultrathin $\mathrm{RuO_{2}}$ films with thicknesses of $t = 1.7$ nm (top) and $t = 9.1$ nm (bottom) \cite{jeong2025metallicity}.}
    \label{fig:Chargetransport}
\end{figure*}
% NOA: b

However, combined ARPES and spin-resolved ARPES (SARPES) studies reported controversial results regarding the existence of spin splitting in $\mathrm{RuO_{2}}$ \cite{ref27,ref28,ref210}. The upper panels of Fig.~\ref{fig:ARPES-AM_NAM} summarize the ARPES and SARPES measurements along the $\Gamma$-$M$ direction reported in Ref.~\cite{ref28}, together with corresponding DFT calculations. The ARPES spectra display a clear band splitting (Fig.~\ref{fig:ARPES-AM_NAM}(a) and two EDC peaks in (c)) and the calculated spin-polarized band structures along two orthogonal directions, $\Gamma_{1}$-$M$-$\Gamma_{2}$ and $\Gamma_{3}$-$M$-$\Gamma_{4}$ (Fig.~\ref{fig:ARPES-AM_NAM}(b)), together with the spin-resolved energy distribution curves at four symmetry-related points (Fig.~\ref{fig:ARPES-AM_NAM}(c)), reveal opposite spin polarizations along the vertical direction. This alternating pattern exhibits the characteristic $d$-wave symmetry expected for the altermagnetic spin texture in $\mathrm{RuO_{2}}$. The lower panels show complementary results from Ref.~\cite{ref27}, which reexamined these features by separating bulk and surface contributions through photon-energy-dependent ARPES. The comparison between single-crystal and thin-film samples (Fig.~\ref{fig:ARPES-AM_NAM}(d)) revealed no $d$-wave like spin splitting. Instead, the SARPES spectra showed an antisymmetric spin polarization distinct from the alternating $d$-wave pattern (Fig.~\ref{fig:ARPES-AM_NAM}(e)), indicating a nonmagnetic origin associated with space-inversion-symmetry-breaking Rashba-like effects. Additional SARPES measurements on the $(100)$, $(110)$, and $(101)$ surfaces of bulk $\mathrm{RuO_{2}}$ \cite{ref210} also detected no spin splitting along any momentum directions. This study emphasizes that the influence of topological surface and interface states must be carefully considered when interpreting the physical and chemical properties of $\mathrm{RuO_{2}}$. %In Ref.~\cite{ref28}, ARPES spectra along $M-\Gamma-M$ momentum cut (upper panels of Fig.~\ref{fig:ARPES-AM_NAM}b) reveal energy band splitting along $\mathit{\Gamma}$-$M$, while those in the lower panels of Fig.~\ref{fig:ARPES-AM_NAM}b depict the spin-resolved bands (red and blue lines representing two opposite spin channels). These results agree with the signature of altermagnetism. However, in Ref.~\cite{ref27}, a similar ARPES analysis performed along Brillouin zone momentum cuts (Fig.~\ref{fig:ARPES-AM_NAM}c) shows the nonmagnetic characteristics in their measured samples. The close resemblance between single-crystal and thin-film band dispersions suggests that intrinsic altermagnetism may not be universally present in $\mathrm{RuO_{2}}$. %In addition, when these ARPES spectra were compared with DFT calculation results, and the experimental results more closely matched the nonmagnetic DFT model as no splitting exceeding $10$ meV was found.
%Moreover, comparison with DFT shows better agreement with the nonmagnetic model, with no splitting exceeding $\sim 10$ meV.

Theoretical analysis suggests that altermagnetic signatures manifest in specific Fermi surfaces, where mirror symmetry-protected magnetic breakdown leads to distinct quantum oscillation frequencies \cite{ref19,Chen_PRB_2025}. These frequencies serve as key experimental markers for identifying the altermagnetic state of $\mathrm{RuO_{2}}$. Nonetheless, bulk $\mathrm{RuO_{2}}$ quantum oscillation measurements show closer agreement with DFT calculations for the nonmagnetic phase \cite{Wu_PRX_2025,qian2025determining}.

\section{Transport properties}

\subsection{Symmetry-tuned charge transport}

Transport measurements provide an effective probe of the intrinsic properties of $\mathrm{RuO_{2}}$. In particular, AHE and thermal transport can reflect symmetry-breaking and Berry-curvature-driven effects characteristic of altermagnets.

Initial transport-based investigations in $\mathrm{RuO_{2}}$ focused on the crystal Hall effect (CHE), a phenomenon first predicted from first-principles calculations and later positioned as a hallmark of altermagnetism. In 2020, Libor Šmejkal {\it{et al.}} predicted a significant crystal-chirality dependent anomalous Hall signal at room temperature in $\mathrm{RuO_{2}}$ \cite{ref40} and named it the CHE. They found that the CHE conductivity depends on the Néel vector orientation and changes sign upon its reversal, reflecting broken crystal symmetry. The sublattices A and B acquire a small canting angle in their magnetic moments due to a weak net magnetic moment $\mathbf{m}$, resulting in weak ferromagnetism. The total Hall conductivity can be written into two components: $\sigma_{xz}=\sigma_{xz}^{\mathrm{AHE}}+ \sigma_{xz}^{\mathrm{CHE}}$, where $\sigma_{xz}^{\mathrm{AHE}}$ is analogous to the AHE in ferromagnets, and $\sigma_{xz}^{\mathrm{CHE}}$ is associated with altermagnetism. As illustrated in Fig. \ref{fig:Chargetransport}(a), for small canting angles, $\sigma_{xz}^{\mathrm{CHE}}$ dominates and is nearly constant, indicating that minor net magnetic moments have little effect on $\sigma_{xz}$ \cite{ref40,ref1}. This symmetry-sensitive CHE prediction has motivated more accessible experimental studies of AHE in $\mathrm{RuO_{2}}$.

To quantify the symmetry-sensitive transport responses, many studies have computed the anomalous Hall conductivity in $\mathrm{RuO_{2}}$ \cite{ref97,ref40,ref54,ref111}. The anomalous Hall conductivity is given by
\begin{equation}
\sigma_{xz}=\frac{e^2}{\hbar} \int \frac{d \mathbf{k}}{(2 \pi)^3} \sum_n f_n(\mathbf{k}) {\Omega}^y_n(\mathbf{k}),
\end{equation}
where ${\Omega}^y_n(\mathbf{k})$ denotes the $y$-component of the Berry curvature of the band $n$ at wavevector $\mathbf{k}$, $f_n(\mathbf{k})$ is the Fermi-Dirac distribution function, $e$ is the elementary charge, and $\hbar$ is the reduced Planck constant. As shown in Fig. \ref{fig:Chargetransport}(b), theoretical calculations reveal that both the magnitude and sign of $\sigma_{xz}$ strongly depend on the altermagnetic exchange field $\bm{J}$, underscoring the crucial role of altermagnetism in the AHE of $\mathrm{RuO_{2}}$ \cite{ref111}.

A key theoretical insight is that the AHE in $\mathrm{RuO_{2}}$ is dictated by crystal symmetry. When the Néel vector aligns along the easy axis $[001]$, symmetry enforces a vanishing Berry curvature, preventing AHE. However, if the Néel vector tilts into the $(001)$ plane, the reduced symmetry allowed for a finite AHE response \cite{ref40,ref45}. To test this symmetry-dependent behavior, Cr doping was introduced in the experiment \cite{ref47}, which serves to rotate the Néel vector from $[001]$ to $[110]$, thereby enabling AHE while preserving collinear antiferromagnetism. However, later DFT study suggests that the observed AHE in Cr-doped $\mathrm{RuO_{2}}$ is from the magnetism of Cr impurities instead of altermagnetism \cite{ref43}. The symmetry constraints on the AHE in altermagnetic materials underscore the critical importance of a tunable means of controlling the Néel vector.

Many transport measurements have provided compelling evidence for the AHE in $\mathrm{RuO_{2}}$ \cite{ref11,ref45,ref71}. Epitaxial $\mathrm{RuO_{2}}$ thin films with various crystallographic orientations exhibit anomalous Hall conductivities exceeding $1000$ $\Omega^{-1}\text{cm}^{-1}$, as revealed by vector magnetometry and transport studies \cite{ref11}. The Hall resistivity is understood as a superposition of three contributions: the ordinary Hall effect (OHE), a field-induced magnetization term (M-AHE), and a term arising from altermagnetic symmetry breaking (L-AHE). Orientation-resolved measurements further clarify this picture: at low temperatures, $\mathrm{RuO_{2}}$ films oriented in $(001)$- and $(100)$-directions exhibit nearly identical field-dependent Hall resistivity curves, attributed to OHE and M-AHE only. In contrast, $(110)$ films display a distinct nonlinear Hall response at high fields (Fig.~\ref{fig:Chargetransport}(c)). As L-AHE is symmetry-forbidden in $(001)$ films, the emergence of nonlinear signals in $(110)$ films serves as strong experimental confirmation of an intrinsic altermagnetic order. These theoretical insights laid the groundwork for symmetry engineering strategies in experimental platforms. The Hall resistance also shows pronounced nonlinear dependence on current and temperature~\cite{ref144}. Two distinct regimes emerge: a low-current plateau and a high-current region where the Hall resistance decays exponentially with increasing current. This threshold behavior depends on the film orientation and is independent of substrate effects. In $(101)$ films, the resistance decreases with increasing temperature below a threshold current, while the critical current for exponential decay increase linearly upon cooling, indicating robust temperature-current scaling. Additionally, external magnetic fields have an exceptionally weak influence on the Hall response, producing a mere $0.001$ $\Omega$ change per 2.8 T, underscoring the stability of $\mathrm{RuO_{2}}$'s antiferromagnetic order against applied magnetic fields. While the fundamental nature of its magnetic ground state remains under debate, these results suggest minimal susceptibility to external field perturbations.

Early studies were constrained by limited magnetic field strengths, leaving open the question of whether the AHE persists at higher fields. Recent high-field experiments by Tschirner {\it et al.} extended the accessible field range and observed saturation of the AHE above 50 T (Fig. \ref{fig:Chargetransport}(d)), thereby reinforcing its intrinsic origin \cite{ref45}. In $(110)$ films, the Hall resistivity exhibits pronounced nonlinearity above 20 T, with the AHE dominating over the OHE across the entire field range. The saturation field $H_c$ is governed by intrinsic interactions, including exchange coupling, magnetocrystalline anisotropy, and the Dzyaloshinskii-Moriya interaction. Furthermore, a subsequent study \cite{jeong2025metallicity} demonstrated that epitaxial strain can stabilize a non-compensated magnetic ground state in $\mathrm{RuO_{2}}$ ultrathin films, leading to the emergence of a sizable AHE even under a moderate magnetic field of $9$ T. As illustrated in Fig.~\ref{fig:Chargetransport}(e), the $t=1.7$ nm film exhibits a prnounced nonlinear Hall response at low temperatures, whereas the thicker ($t=9.1$) nm film shows an almost perfectly linear behavior across the entire temperature range. This observation provides compelling evidence that strain engineering offers an effective means of inducing and modulating the intrinsic magnetism in $\mathrm{RuO_{2}}$ thin films.

Complementing these findings, the planar Hall effect \cite{ref206,ref71}, in-plane Hall effect (IPHE) \cite{ref142,ref207} and high-order nonlinear Hall effect \cite{Chu2025nonlinear} serve as a symmetry-sensitive transport signature that is distinct from the AHE. While AHE results from relativistic magnetic symmetry breaking, the IPHE arises due to reduced crystallographic symmetry within specific planes. Experimentally, a significant $B$-linear IPHE has been observed in $(111)$ and $(101)$ $\mathrm{RuO_{2}}$ thin films but is absent in $(100)$ and $(001)$ orientations, confirming its strong dependence on crystal symmetry \cite{ref142}. In addition, rencently, a third-order nonlinear Hall effect is observed in $(001)$ $\mathrm{RuO_{2}}$ thin films with a twofold angular dependence, exhibiting the transport fingerprint of altermagnets \cite{Chu2025nonlinear}. The emergence of IPHE and high-order nonlinear Hall effect in planes with specific symmety introduces a new axis of anisotropy in $\mathrm{RuO_{2}}$ transport studies, highlighting the rich interplay between crystallographic orientation and Hall responses. 

Nevertheless, a recent study reported that the magnetoresistance signals in the $[001]$ and $[010]$ directions are comparable, suggesting that the expected strong orientation dependence has not yet been observed experimentally~\cite{akashdeep2025interfacegenerated}. Although this finding dose not provide a definitive conclusion regarding whether $\mathrm{RuO_{2}}$ is an altermagnet, it clearly indicates that no transport signal attributable to the intrinsic spin polarization associated with altermagnetism was detected under the present measurement conditions.

%A recent study on the electrical transport properties of $\mathrm{RuO_{2}}$ demonstrated that its charge dynamics are primarily governed by the inverse spin Hall effect (ISHE) and electrical anisotropic conductivity (EAC), with pronounced orientation dependence \cite{ref199}. The combined influence of ISHE and EAC appears only in the $(101)$ orientation, whereas the $(001)$ orientation exhibits no detectable contribution from EAC and is solely characterized by ISHE. Moreover, THz is attributed exclusively to ISHE in all orientations. Crucially, the study found no evidence of the inverse altermagnetic spin-splitting effect (IASSE), which is typically regarded as a key signature of altermagnetism. This finding further challenges the presence of intrinsic altermagnetism in $\mathrm{RuO_{2}}$. 

%However, a spin-splitting magnetoresistance (SSMR) effect has been proposed, exhibiting distinct angular and temperature dependencies compared to conventional spin Hall magnetoresistance (SMR). Magnetotransport measurements on $(101)$-oriented $\mathrm{RuO_{2}}$/Co bilayers reveal that the observed magnetoresistance comprises contributions from both SSMR and SMR. The emergence of SSMR signals the presence of altermagnetic order in the $\mathrm{RuO_{2}}$ layer. Symmetry analysis further identifies the Néel vector orientation along the $[001]$ axis, corroborating the intrinsic altermagnetism of $\mathrm{RuO_{2}}$. Together, these orientation-dependent phenomena reveal a complex and tunable transport landscape in $\mathrm{RuO_{2}}$ thin films.

\subsection{Thermal transport}

\begin{figure*}[t]
    \centering
    \includegraphics[width=1.0\linewidth]{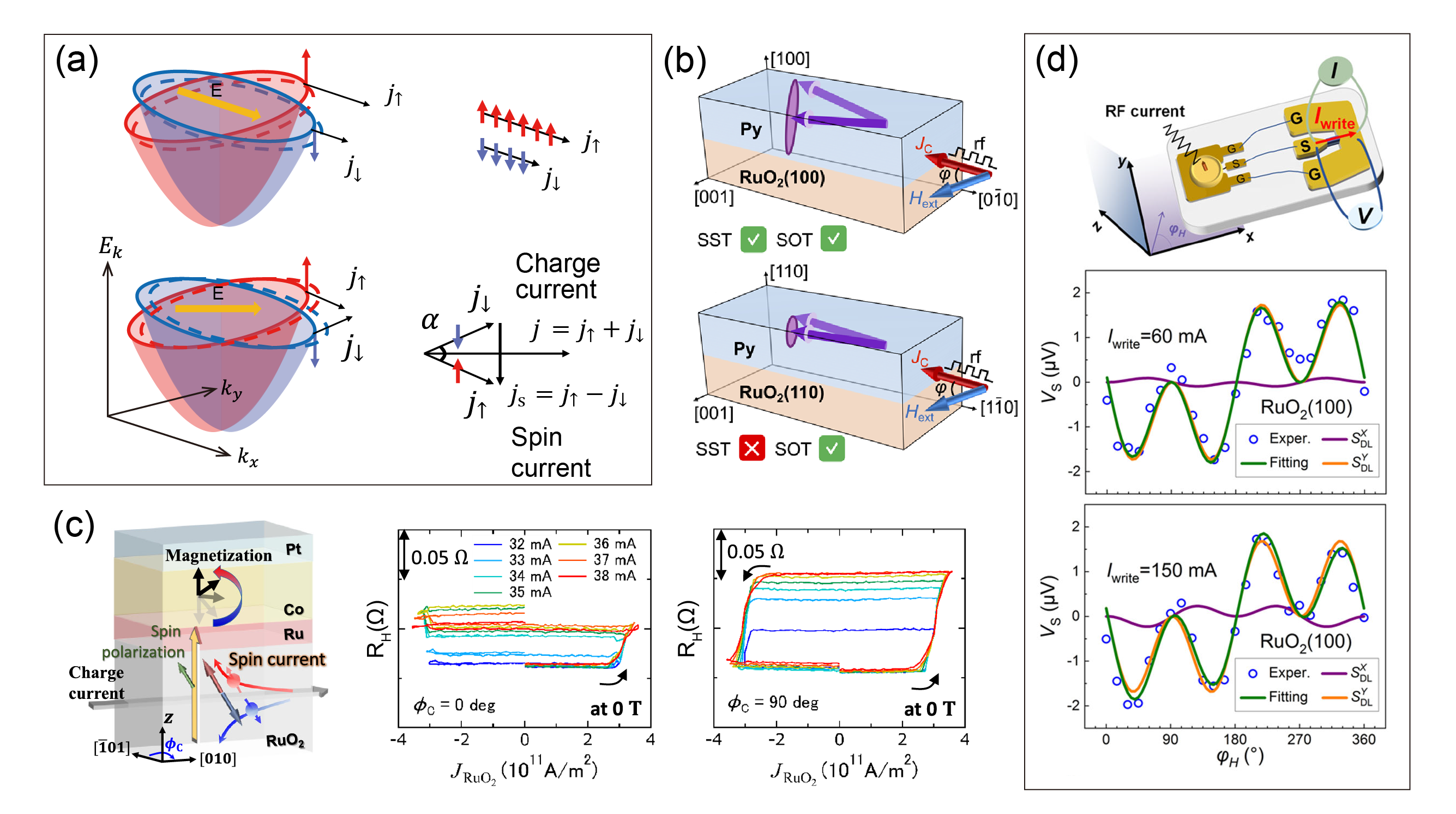}
    \caption{ Spin splitting torque and orientation-dependent spin transport in $\mathrm{RuO_{2}}$. (a) Schematic of spin splitting effect in $\mathrm{RuO_{2}}$. (b) Spin splitting torque dependent on crystal symmetry \cite{ref3}. For the $\mathrm{RuO_{2}}$ $(100)$/Py and $\mathrm{RuO_{2}}$ $(110)$/Py samples. (c) Schematic illustration of z-polarized spin current injection into Co layer by spin splitting torque \cite{ref21}, and Hall resistance of the Co layer along the charge current at $\phi_c=0^{\circ}$ and $\phi_c=90^{\circ}$. (d) Schematic of the ST-FMR measurement and the spin splitting torque signals measured in Pt/$\mathrm{RuO_{2}}$ $(100)$/Py device under different writing currents. The spin splitting torque voltage $V_{S}$ as a function of the in-plane magnetic-field angle $\varphi_{H}$ is shown for $I_{\mathrm{write}}=60$ mA (middle) and $150$ mA (bottom) \cite{ref200}.}
    \label{fig:Spin current-SST}
\end{figure*}
% NOA: b,c

The Néel-vector-orientation dependent crystal Nernst effect and crystal thermal Hall effect were further predicted in $\mathrm{RuO_{2}}$ as the thermoelectric and thermal analogs of the crystal Hall effect~\cite{ref50}. The anisotropy of thermal transport coefficients highlights the interplay between spin and crystal symmetry in altermagnets. Additionally, theoretical modeling of an insulating altermagnet suggests that thermal conductivity remains finite, strongly depends on the Néel vector, and is tunable via strain \cite{ref48,ref18}. Complementary theoretical work based on the Boltzmann transport equation further reveals that $\mathrm{RuO_{2}}$ can generate a pronounced magnon spin Nernst effect, whose magnitude is comparable to, or even exceeds, the electronic contribution. In comparison with MnTe, the origin of this substantial spin-thermal response in $\mathrm{RuO_{2}}$ is attributed to the characteristic planar magnon splitting \cite{wu2025magnon}. These predictions emphasize the potential of $\mathrm{RuO_{2}}$ as a platform for exploring altermagnetic transport phenomena beyond conventional charge-based effects.

Measurements on $\mathrm{RuO_{2}}$ single crystals reveal that the Seebeck coefficient exhibits a smooth and continuous increase between 300 K and 970 K, showing no features associated with the onset of magnetic ordering \cite{ref60}. The measured electrical resistivity also increases monotonically from low temperature up to 970 K, without any slope change or anomaly that would indicate a magnetic phase transition. These high-temperature transport characteristics are therefore inconsistent with the existence of long-range magnetic order in bulk $\mathrm{RuO_{2}}$. 

In summary, from the theoretically predicted CHE to experimentally observed various Hall responses, these transport signatures consistently reflect the role of Néel vector orientation and crystallographic symmetry in governing conductivity, thus supporting $\mathrm{RuO_{2}}$ as an altermagnet. However, the magnetoresistance and Seebeck coefficient measured in bulk $\mathrm{RuO_{2}}$ exhibit no signiture of altermagnetism.

\subsection{Spin splitting torque effect}

The spin splitting effect in $\mathrm{RuO_{2}}$ is a fundamentally distinct mechanism from the AHE, differing in both physical origin and functional application. While the AHE is a relativistic phenomenon linked to charge current responses mediated by SOC and magnetic order, the spin splitting in $\mathrm{RuO_{2}}$ arises from a non-relativistic spin current response. This mechanism enables direct electrical control over magnetic order and reflects the symmetry-protected nature of altermagnetic spin transport.

\begin{figure*}[t]
    \centering
    \includegraphics[width=1.0\linewidth]{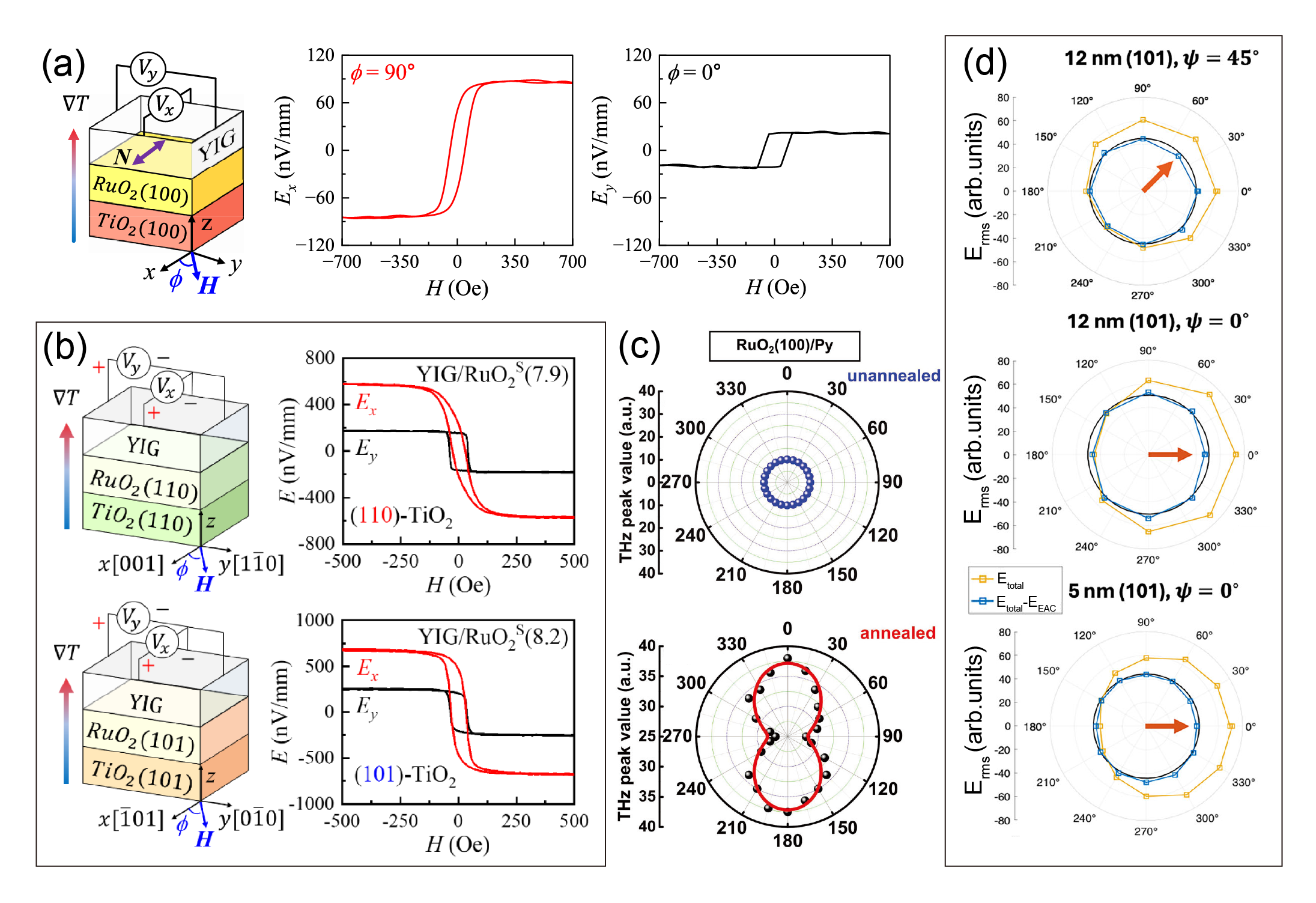}
    \caption{ Experimental signatures supporting and excluding the IASSE in $\mathrm{RuO_{2}}$. (a) Schematic of the experiment setup for the $(100)$-oriented YIG/$\mathrm{RuO_{2}}$/$\mathrm{TiO_{2}}$ sample and the measured spin Seebeck voltage signals. The spin-to-charge conversion electric fields $E_x$ and $E_y$ are detected along the $x$ and $y$ axes, with $\phi_H$ denoting the angle between the external magnetic field and the $x$-axis \cite{ref26}. (b) Experiment setups and spin Seebeck voltage measurements for $(110)$- and $(101)$-oriented YIG/$\mathrm{RuO_{2}}$/$\mathrm{TiO_{2}}$ samples. Both orientations exhibit angular anisotropy in the detected voltages \cite{wangrobust}. (c) Angular dependence of the THz emission amplitude for the unannealed and annealed $\mathrm{RuO_{2}}(100)$/Py bilayers. Before annealing, the THz signal is nearly isotropic, consistent with a dominant contribution from the crystal-independent inverse spin Hall effect. After magnetic-field annealing, the THz emission develops a pronounced two-fold anisotropy, suggesting a possible contribution from the IASSE \cite{ref100}. (d) Polar plots of THz emission amplitude ($E_{\mathrm{rms}}$) as a function of the external magnetic field angle for the 12 nm and 5 nm $\mathrm{RuO_{2}}$$(101)$ film. Yellow and blue curves denote the total emission ($E_{\mathrm{total}}$) and the inverse spin Hall effect component after removing the electric-field-induced artifact ($E_{\mathrm{total}} - E_{\mathrm{EAC}}$), respectively. The red arrows mark the electrical anisotropic conductivity (EAC) contribution direction, which rotates with the crystal azimuthal angle $\psi$. No IASSE-related anisotropy is observed \cite{ref199}.}
    \label{fig:IASSE}
\end{figure*}
% NOA: e,f

Theoretical studies have uncovered a unique spin current generation mechanism in $\mathrm{RuO_{2}}$, rooted in its crystal field environment and magnetic symmetry \cite{ref13}. First-principles calculations reveal anisotropic spin-split bands, independent of SOC, arising from local Ru-O coordination \cite{ref13,ref40}. This mechanism is highly directional, symmetry-protected, and robust even at room temperature. As illustrated in the Fig.~\ref{fig:Spin current-SST}(a): (i) With an electric field applied along $[110]$, the charge currents of spin-up and spin-down channels flow in the same direction but with different magnitudes, generating a net spin current aligned with the electric field; (ii) For an electric field along $[100]$, the charge currents with spin-up and spin-down channels are deflected in directions at an angle of about $34^\circ$, resulting in a spin current perpendicular to the field. The latter case underpins the spin splitting torque (SST) mechanism, in which an in-plane current in $\mathrm{RuO_{2}}$ injects a spin current into an adjacent ferromagnet along the out-of-plane direction~\cite{ref13}. When the Néel vector lies along the $[001]$ easy axis, this torque can efficiently switch the magnetic state. %Additionally, the Néel vector in $\mathrm{RuO_{2}}$ can be electrically switched via spin orbit torque (SOT) from the underlying Pt layer, such SOT-driven reorientation directly enables the control of both the magnitude and polarization of the SST~\cite{ref200}. As shown in Fig.~\ref{fig:Spin current-SST}(d), increasing the writing current from 60 mA to 150 mA switches the Néel vector toward the $x$ axis, which selectively enhances the $x$-polarized damping-like torque component $S^{X}_{\mathrm{DL}}$ (purple), whereas the $y$-polarized component $S^{Y}_{\mathrm{DL}}$ (orange) remains largely unchanged. This clearly demonstrates that manipulating the Néel-vector orientation allows directional engineering of SST in altermagnetic $\mathrm{RuO_{2}}$.In $\mathrm{RuO_{2}}$ $(100)$ films, spin currents originate from both the spin splitting and spin Hall effect, whereas in $(110)$ films, only the spin Hall effect is allowed by symmetry.
Notably, SST integrates the advantages of spin transfer torque (STT) and spin orbit torque (SOT), while mitigating their respective limitations such as out-of-plane current injection and spin loss \cite{ref3}.

Experimental validation of these predictions has been achieved using spin–torque ferromagnetic resonance (ST-FMR) measurements in $\mathrm{RuO_{2}}$-based heterostuctures \cite{ref5,ref3,ref21,ref75,ref15,ref200}. Orientation-resolved studies on $\mathrm{RuO_{2}}(100)$/Py and $\mathrm{RuO_{2}}(110)$/Py demonstrated that spin current generation is highly dependent on crystal symmetry. In $\mathrm{RuO_{2}}(100)$/Py, a charge current along the $[010]$ axis produces an out-of-plane spin current along the $[100]$ axis (Fig.~\ref{fig:Spin current-SST}(b)), confirming the coexistence of SST and SOT. In contrast, $\mathrm{RuO_{2}}(110)$/Py exhibits symmetry-imposed suppression of SST while still allowing SOT. These findings are further corroborated by magnetization switching experiments in Co layers \cite{ref21}. When a charge current is injected along the $[101]$ direction ($\phi_c=0^{\circ}$), no out-of-plane spin current is generated and magnetic switching is prevented (Fig.~\ref{fig:Spin current-SST}(c)). Conversely, when the charge current is applied along the $[010]$ axis ($\phi_c=90^{\circ}$), a strong out-of-plane spin current enables magnetic switching with up to $75\%$ efficiency. 

Additionally, the Néel vector in $\mathrm{RuO_{2}}$ can be electrically switched via spin orbit torque (SOT) from the underlying Pt layer, such SOT-driven reorientation directly enables the control of both the magnitude and polarization of the SST~\cite{ref200}. As shown in Fig.~\ref{fig:Spin current-SST}(d), increasing the writing current from 60 mA to 150 mA switches the Néel vector toward the $x$ axis, which selectively enhances the $x$-polarized damping-like torque component $S^{X}_{\mathrm{DL}}$ (purple), whereas the $y$-polarized component $S^{Y}_{\mathrm{DL}}$ (orange) remains largely unchanged. This clearly demonstrates that manipulating the Néel-vector orientation allows directional engineering of SST in altermagnetic $\mathrm{RuO_{2}}$.In $\mathrm{RuO_{2}}$ $(100)$ films, spin currents originate from both the spin splitting and spin Hall effect, whereas in $(110)$ films, only the spin Hall effect is allowed by symmetry.

Further progress in spin current generation has been achieved by engineering $\mathrm{RuO_{2}(101)/[Co/Pt]_{2}/Ta}$ heterostructures through controlled magnetic sputtering~\cite{ref110}. These films demonstrated nearly $100 \%$ field-free SOT switching. The SOT switching measurements reveal that: (1) at $\theta = 90^{\circ}$ (corresponding to a current applied along the $[010]$ axis), near complete switching occurs without any external field; (2) at $\theta = 0^{\circ}$, switching is absent. Magnetic Kerr microscopy tracked domain evolution throughout the process, offering real-time confirmation of full, field-free SOT switching. These results not only support the non-relativistic spin splitting in the band structure of $\mathrm{RuO_{2}}$ but also demonstrate its practical utility in device-level magnetic control.

\subsection{Inverse altermagnetic spin-splitting effect}

To explore the spin-charge conversion in $\mathrm{RuO_{2}}$, several experiments examined the inverse altermagnetic spin-splitting effect (IASSE). Spin Seebeck effect measurements were first carried on $\mathrm{RuO_{2}}$-based heterostructures, which revealed that the spin-charge conversion efficiency strongly depends on the crystal axis and decreases markedly with increasing temperature \cite{ref2}. This trend contrasts with other nonmagnetic materials such as heavy metal Pt and rutile $\mathrm{IrO_{2}}$, where the conversion efficiency is largely insensitive to temperature \cite{ref2}. In $\mathrm{RuO_{2}}$, the enhanced efficiency at low temperatures is attributed to enhanced electron lifetimes, which amplify the spin transport signal. Since the spin-charge conversion based on the nonrelativistic altermagnetic spin splitting effect is related to a longitudinal transport mechanism, it is thus dependent on the scattering. On the contrary, the spin-charge conversion based on the spin Hall effect is analogous to the anomalous Hall transport, which can be described by the language of scattering-independent spin Berry curvature. Therefore, the observed temperature-dependent spin-charge conversion in $\mathrm{RuO_{2}}$ reflects the presence of nonrelativistic, time-reversal-odd altermagnetic spin splitting~\cite{ref13,ref5}, which differs fundamentally from conventional, time-reversal-even spin-orbit coupling effects. Subsequent studies using spin Seebeck effect have successfully separated the contributions from the IASSE and the inverse spin Hall effect (ISHE) in $\mathrm{RuO_{2}}$ \cite{ref26}. Spin Seebeck measurements on $(100)$-oriented YIG/$\mathrm{RuO_{2}}$/$\mathrm{TiO_{2}}$ sample (Fig.~\ref{fig:IASSE}(a)) demonstrated that the voltage component measured along the $y$-axis contains both ISHE and IASSE contributions when the injected spin polarization aligns with the Néel vector, whereas the $x$-axis voltage only reflects the ISHE when the spin polarization is perpendicular to Néel vector.

Compared with conventional spin-source materials, the non-relativistic altermagnetic spin splitting effect (ASSE) in $\mathrm{RuO_{2}}$ enables efficient charge-spin conversion without relying on strong SOC \cite{ref75}. Notably, $\mathrm{RuO_{2}}$ can produce T-even spin currents through SHE \cite{ref88} and T-odd spin currents via the ASSE \cite{ref59,ref13,ref143}, a distinction that is critical for the development of high-performance spintronic devices. Despite the absence of strong SOC, ST-FMR and spin pumping experiments reveal that $\mathrm{RuO_{2}}$ $(100)$ thin films exhibit a large spin Hall angle of $0.183$ and a spin diffusion length exceeding $12$ nm, surpassing those of typical heavy metals~\cite{ref75}. In addition, $\mathrm{RuO_{2}}$ (101) films generate spin currents with out-of-plane spin polarization while maintaining a substantial spin diffusion length of $11.0 \pm 2.1$ nm. The SOC-independent nature of the ASSE minimizes spin relaxation, thereby enhancing both spin diffusion and charge-spin conversion efficiency. These features render $\mathrm{RuO_{2}}$ a promising candidate for low-power spintronic applications. The experimental result underscores the critical role of crystallographic orientation in shaping spin transport behavior.

Early THz-emission measurements on $\mathrm{RuO_{2}}(100)$/Py bilayers reported that annealing greatly enhances the THz amplitude and induces pronounced anisotropy, suggesting successful alignment of the Néel vector and a potential IASSE contribution \cite{ref100} (Fig.~\ref{fig:IASSE}(c)). In contrast, $\mathrm{RuO_{2}}(110)$/Py samples exhibit strictly isotropic THz emission regardless of annealing, consistent with symmetry arguments that forbid IASSE in the $(110)$ orientation and thereby confirming that the anisotropy observed in the (100) case originates from the crystal-dependent mechanism associated with the proposed altermagnetic order \cite{ref100}. 

Recently, spin pumping and ST-FMR measurements demonstrated that the ISHE provides the predominant component of the spin-charge conversion signal \cite{ref118}. More recent work has reinforced this conclusion. Spin Seebeck effect \cite{wangrobust} and THz emission measurements \cite{jechumtalspintochargecurrent} independently found that the observed anisotropy signal is fully accounted for the anisotropic spin Hall effect. Spin Seebeck effect measurements on $(110)$- and $(101)$-oriented YIG/$\mathrm{RuO_{2}}$/$\mathrm{TiO_{2}}$ samples (Fig.~\ref{fig:IASSE}(b)) show similar angular anisotropy in the detected voltages, because the $(110)$ symmetry strictly forbids any IASSE contribution, the anisotropy must arise entirely from the anisotropic SHE \cite{wangrobust}. THz measurements reported nearly identical anisotropic responses in $\mathrm{RuO_{2}}$ $(100)$ and $(110)$ films, clearly indicating the absence of any measurable IASSE~\cite{jechumtalspintochargecurrent}. More recently, comprehensive THz measurements (Fig.~\ref{fig:IASSE}(d)) across $(100)$,$(110)$ and $(101)$ oriented samples showed that the angular dependence of the emitted THz field can be quantitatively explained by electrical anisotropic conductivity, which rotates with the crystal angle \cite{ref199}. Taken together, these results indicate that no experiment to date has yielded definitive evidence of IASSE in $\mathrm{RuO_{2}}$, despite the effect being considered a key transport signature of altermagnetism. 

In summary, although numerous reports support the altermagnetic spin splitting effect in $\mathrm{RuO_{2}}$ through various IASSE measurements, recent studies have challenged these findings, questioning whether the IASSE signal can be effectively separated from the inverse spin Hall effect. Nevertheless, the high efficiency of spin-charge conversion in $\mathrm{RuO_{2}}$-based devices remains undisputed. The discrepancies in experimental reports likely arise from interface effects involving adjacent materials. Theoretical work using a bilayer model, parameterized by first-principles calculations, demonstrates that even when altermagnetic order is present at the interface, its interplay with interface-induced Rashba spin-orbit coupling can generate highly efficient nonlinear spin–charge conversion via an anomalous Edelstein mechanism~\cite{trama2025nonlinear}. Furthermore, recent experiments have shown that the adjacent ferromagnet in $\mathrm{RuO_{2}}$ heterostructures can dictate both the magnitude and the sign of the spin-charge conversion~\cite{yang2025ferromagnetic}.

\subsection{Altermagnetic tunnel junctions}

Given its efficient spin-charge conversion and anisotropic transport behavior, $\mathrm{RuO_{2}}$ is also a strong candidate for magnetic tunnel junctions (MTJs). DFT and transport calculations for $\mathrm{RuO_{2}}$/$\mathrm{TiO_{2}}$/$\mathrm{RuO_{2}}$$(001)$ MTJs predicted a remarkable tunneling magnetoresistance (TMR) ratio up to $500\%$ at the Fermi level \cite{ref38}, arising from spin-polarized conduction channels sensitive to the alignment of Néel vectors. A similar study on $\mathrm{RuO_{2}}$/$\mathrm{TiO_{2}}$/$\mathrm{RuO_{2}}$$(110)$ MTJs revealed both giant TMR and significant STT \cite{ref20}. These effects originate from the spin-split Fermi surface and momentum-dependent spin polarization intrinsic to altermagnets. Notably, device performance also exhibits strong crystallographic anisotropy. For instance, for $\mathrm{RuO_{2}}$$(001)$/$\mathrm{TiO_{2}}$/$\mathrm{CrO_{2}}$ and $\mathrm{RuO_{2}}$$(110)$/$\mathrm{TiO_{2}}$/$\mathrm{CrO_{2}}$ MTJs, a pronounced directional dependence is observed \cite{ref7,ref35}. Such pronounced directionality further supports the altermagnetic origin of spin splitting in $\mathrm{RuO_{2}}$. Despite this promise, early work pointed to a low density of states near the Fermi level in $\mathrm{RuO_{2}}$, limiting TMR contributions \cite{ref141}. To address this, doping strategies such as substitution in $\mathrm{Ru}_{1-x}\mathrm{Cr}_{x}\mathrm{O}_{2}$ ($0.3\leq x\leq0.5$) have been employed. These enhance both magnetic moment and electronic correlations, thereby strengthening spin polarization critical to tunneling behavior. Local density of states trends match observed transport characteristics, shedding light on the microscopic origins of TMR enhancement \cite{ref44}. A very recent study \cite{xu2025giant} on all-epitaxial $\mathrm{RuO_{2}}$/$\mathrm{MgO}$/$\mathrm{RuO_{2}}$ antiferromagnetic tunnel junctions demonstrated a tunneling anisotropic magnetoresistance (TAMR) ratio as high as approximately $60 \%$  achieved through a spin-flop mechanism. At zero or low magnetic fields, the Néel vectors of the top and bottom $\mathrm{RuO_{2}}$ electrodes are aligned parallel to each other within the film plane, resulting in a low tunneling resistance. Upon applying a sufficiently strong in-plane magnetic field of about $6$ T, the magnetic moments in the thinner bottom $\mathrm{RuO_{2}}$ layer undergo a spin-flop transition, making the Néel vectors of the two electrodes mutually perpendicular and thereby producing a large TAMR signal. Moreover, the TAMR magnitude exhibits a pronounced dependence on the angle $\theta$ between the applied magnetic field and the crystallographic axis of $\mathrm{RuO_{2}}$, following a $\sin^2{\theta}$ functional relationship. This angular dependence reveals a pronounced in-plane magnetocrystalline anisotropy. As the temperature decreases from $300$ K to $3$ K, the TAMR ratio increases from $60 \%$ to $73 \%$, suggesting that the antiferromagnetic order in $\mathrm{RuO_{2}}$ becomes more robust at lower temperatures, thereby enhancing the magnetoresistive effect.

\section{Other aspects}

\subsection{Superconducting properties}

Epitaxial strain in $\mathrm{RuO_{2}}$ thin films not only allows tuning of the intrinsic altermagnetic order but can also induce superconductivity. Moreover, when integrated with conventional superconductors to form hybrid junctions, it gives rise to rich superconducting transport phenomena, such as unconventional Josephson effects.

\begin{figure*}[t]
    \centering
    \includegraphics[width=1.0\linewidth]{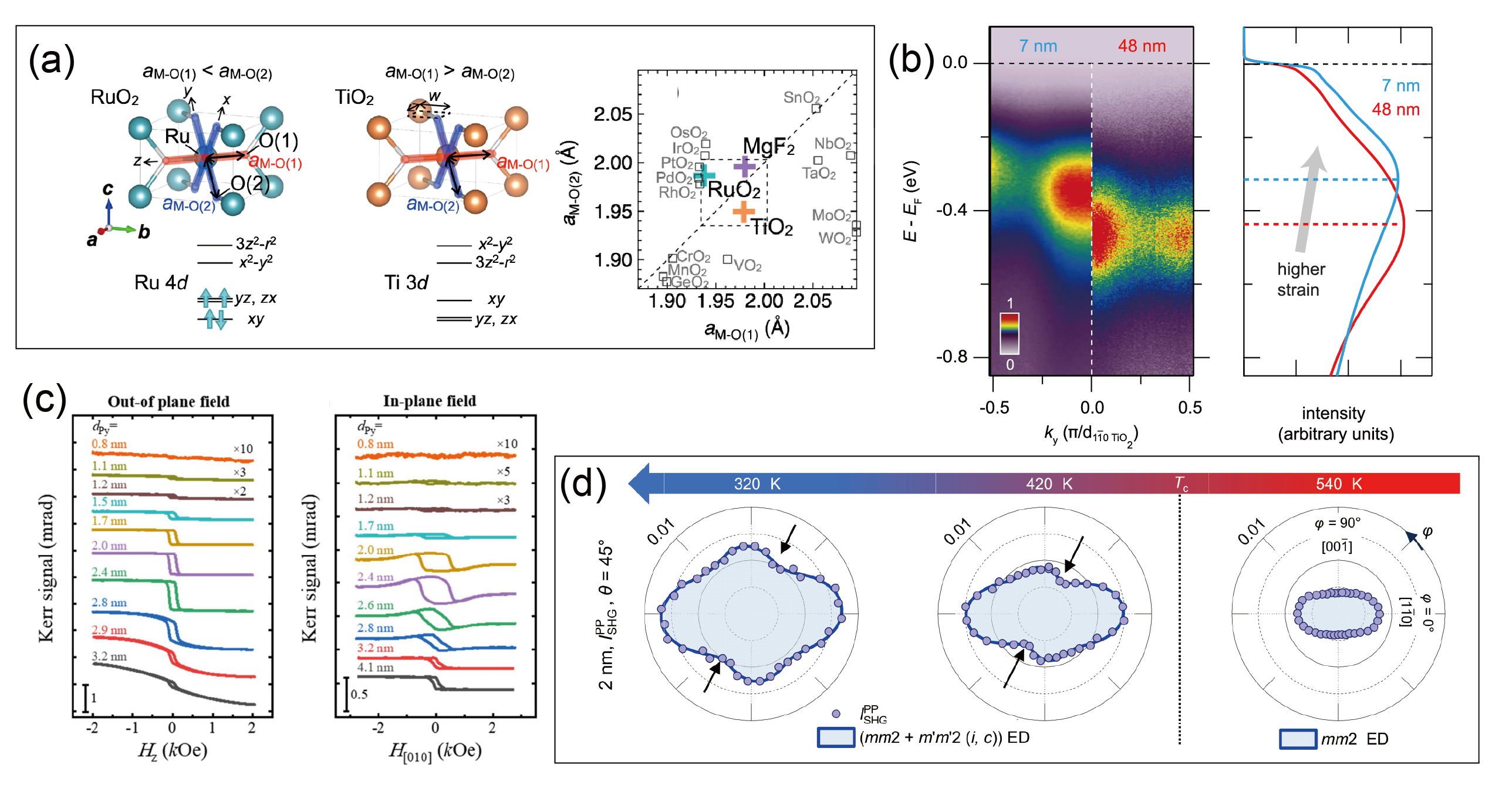}
    \caption{ Strain-driven symmetry breaking and its electronic and optical manifestations in $\mathrm{RuO_{2}}$. (a) $\mathrm{RuO_{2}}$ rutile crystal structure composed of short $\mathrm{\mathrm{M\text{-}O(1)}}$ and long $\mathrm{M\text{-}O(2)}$, and the $\mathrm{TiO_{2}}$ composed of long $\mathrm{M\text{-}O(1)}$ and short $\mathrm{M\text{-}O(2)}$, as well as the mapping of the two bond lengths $a_{\mathrm{M\text{-}O(1)}}$ vs. $a_{\mathrm{M\text{-}O(2)}}$ obtained based on the lattice parameters $a$ and $c$ of many rutile compounds and the Wyckoff position coordinates \cite{ref46}. (b) ARPES spectra of highly strained ($7$ nm) and partially relaxed ($48$ nm) $\mathrm{RuO_{2}}$ $(110)$ thick films, together with the corresponding energy distribution curves obtained by integrating the ARPES intensity \cite{ref33}. (c) Kerr hysteresis loops of $\mathrm{Al_{2}O_{3}}$/Py/$\mathrm{RuO_{2}}$/$\mathrm{Al_{2}O_{3}}$(sub.) heterostructures with varying Py thickness $d_{\mathrm{Py}}$, measured under out-of-plane and in-plane magnetic fields \cite{ref145}. (d) Temperature-dependent SHG polar patterns of a 2 nm fully strained $\mathrm{RuO_{2}}$ film. Below the transition temperature ($\sim 500$ K), the SHG signal shows characteristic asymmetric indentations (arrows), indicating a symmetry-broken low-temperature phase. Above 500 K, the pattern becomes symmetric, consistent with the high-symmetry nonmagnetic phase \cite{ref70}.}
    \label{fig:SC-optical}
\end{figure*}
%NOA: a,d

Resistivity measurements on $\mathrm{RuO_{2}}$ films grown epitaxially on (110)-oriented $\mathrm{TiO_{2}}$ and $\mathrm{MgF_{2}}$ substrates reveal a striking contrast: superconductivity emerges in films on $\mathrm{TiO_{2}}$, whereas those on $\mathrm{MgF_{2}}$ remain metallic \cite{ref46}. First-principles calculations attribute this substrate-dependent behavior to a significant reduction in the metal-oxygen bond length $a_\mathrm{{M\text{-}O(2)}}$ in superconducting films, implying that c-axis compression or $\mathrm{{M\text{-}O(2)}}$ bond shortening is a key structural parameter for the onset of superconductivity (Fig.~\ref{fig:SC-optical}(a)). These results highlight the critical role of epitaxial strain in modulating the electronic ground states of $\mathrm{RuO_{2}}$, providing a strain-induced pathway for the metal-to-superconductor transition \cite{ref46}. Notably, the experimental observations align well with theoretical predictions regarding the tunability of superconductivity under lattice distortion \cite{ref33,wadehra2025straininduced}. As shown in Fig.~\ref{fig:SC-optical}(b), the energy distribution curves extracted from the ARPES spectra reveal a substantially enhanced spectral weight near the Fermi level in the highly strained $7$ nm film compared with the partially relaxed $48$ nm film. This strain-induced increase in the low-energy density of states closely follows the trend predicted by DFT calculations.

Beyond its influence on superconductivity, strain also modulates the magnetic and altermagnetic properties of $\mathrm{RuO_{2}}$~\cite{ref46}. Strain-induced effects, such as the azimuthal modulation of the $110$ Bragg peak intensity and the symmetry lifting of structural sublattices, have been shown to enhance the density of states at the Fermi level. These structural and electronic modifications affects not only the magnetization direction but also the emergence and tunability of the altermagnetic order~\cite{ref14,ref30}.

While strain-induced superconductivity concerns intrinsic properties of $\mathrm{RuO_{2}}$, another line of research examines how superconducting transport behaves when $\mathrm{RuO_{2}}$ functions instead as non-superconducting altermagnetic metallic layers in hybrid superconducting junctions, most notably Josephson junctions~\cite{ref133,ref139,ref132}. These setups do not rely on superconductivity within $\mathrm{RuO_{2}}$ itself. Rather, they explore the general principles of how an altermagnetic barrier modifies the Josephson effect when placed between two conventional superconductors. Model calculations for junctions containing an altermagnetic spacer reveal oscillatory supercurrents around zero when varying the spacer length, chemical potential in the altermagnet, or junction orientation, which is distinct from both ferromagnetic and antiferromagnetic counterparts~\cite{ref132,ref133}. %Ferromagnetic junctions usually show oscillations with strong exponential decay due to a finite magnetic moment, while antiferromagnetic junctions display a $\pi$ state only for an odd number of atomic layers. In contrast, %Altermagnetic junctions display weakly damped oscillations in the supercurrent when changing the junction length.  
Crucially, the oscillation pattern occurs in the absence of a net mangetization and depends strongly on the junction orientation relative to the crystallographic axes of the altermagnetic layer. Orientations for which the Fermi-surface spin splitting vanishes in the corresponding momentum direction result in the largest oscillation period. These oscillations correspond to a sequence of $0$-$\pi$ transitions of the junction~\cite{ref132,ref133} and stem from the emergence of finite-momentum Cooper pairs across the altermagnet~\cite{ref132}. Interestingly, when the junction is oriented along directions with vanishing spin splitting, the dominant Cooper-pair trajectories propagate at a large oblique angle relative to the transport direction~\cite{ref132}. As a result, the critical supercurrent becomes highly sensitive to the junction aspect ratio and develops an anomalous Fraunhofer pattern under perpendicular magnetic fields. Theoretical modeling using parameters characteristic of $\mathrm{RuO_{2}}$ indicates that the oscillations can be observed in junctions with lengths exceeding twenty nanometers~\cite{ref132}, making them experimentally accessible. It should be emphasized, however, that the existing studies are based on idealized altermagnetic model Hamiltonians rather than the material-specific electronic structure of $\mathrm{RuO_{2}}$. Thus, their predictions only illustrate the generic behavior expected for an altermagnetic barrier. A realistic description of $\mathrm{RuO_{2}}$-based junctions will require future calculations that incorporate the actual electronic structure of the material.

These unique transport and spectroscopic features have spurred growing interest in the transmission characteristics of superconductor–altermagnet or superconductor–altermagnet–superconductor junctions \cite{ref137,ref138,ref139,ref133,ref132,WMiao24PRB,QCheng24PRB,QCheng24PRB-b,ref140,HPSun25PRB,Nagae25PRB}. The altermagnetic order in $\mathrm{RuO_{2}}$ imparts finite center-of-mass momentum to Cooper pairs despite the zero net magnetization in the system \cite{ref132}, leading to the formation of $\pi$-state and unconventional current-phase relations in Josephson junctions. These effects are robust against variations in the details of the $s$-wave superconductors and the altermagnet–superconductor interfaces. Notably, unlike in ferromagnetic or antiferromagnetic Josephson junctions where a finite net magnetization is required \cite{ref134,ref201,ref202,ref136,ref135,ref204,ref203}, the $\pi$ junction state and $0$–$\pi$ transitions in altermagnetic junctions occur without net magnetism. These findings open new opportunities for integrating altermagnetic materials into unconventional superconducting platforms, with potential applications in topological quantum devices \cite{ref140}.

\subsection{Optical properties}

Beyond transport and superconducting behavior, the interplay between crystal symmetry and magnetic ordering in $\mathrm{RuO_{2}}$ gives rise to unique magneto-optical effects, offering additional insights into its altermagnetic nature \cite{ref119,ref145,ref146}. Recent studies have discovered unconventional magneto-optical effects in certain antiferromagnets, which occur independently of conventional mechanisms and may be attributed to crystal chirality associated with nonmagnetic atoms \cite{ref120,ref126,ref127,ref123,ref130,ref131,ref121,ref122,ref124,ref125,ref128,ref129}. In $\mathrm{RuO_{2}}$, first-principles calculations demonstrate that the signs of Kerr and Faraday spectra reverse upon inversion of crystal chirality. Moreover, their magnitudes are tunable through reorientation of the Néel vector \cite{ref119}. These findings indicate a strong coupling between lattice symmetry and magnetic order, enabling optical detection of both structural chirality and spin configuration. Remarkably, the Kerr rotation angle in $\mathrm{RuO_{2}}$ can reach up to $0.62^{\circ}$, exceeding that observed in many traditional ferromagnets \cite{ref119}. The spectral integrals of the Kerr and Faraday effects exhibit a strong correlation with the magnetocrystalline anisotropy energy and orbital magnetic moments, suggesting that these magneto-optical responses may serve as an effective optical probe for detecting the crystal structure chirality and Néel vector orientation. %Further theoretical analysis \cite{ref48} of $\mathrm{RuO_{2}}$'s magnon spectrum reveals chiral splitting of magnon bands, analogous to the spin splitting observed in its electronic band structure, both exhibiting alternating variations across the Brillouin zone. In contrast to the quadratic dispersion observed in ferromagnets, most regions of the magnon spectrum of $\mathrm{RuO_{2}}$ display linear dispersion.

Experimental observations also support the presence of magneto-optical phenomena at $\mathrm{RuO_{2}}$ interfaces. In particular, magneto-optical Kerr effect (MOKE) measurements of permalloy (Py) thin films deposited on $\mathrm{RuO_{2}}$ $(101)$ films reveal perpendicular magnetic anisotropy in the $1.0-2.5$ nm thickness range (Fig.~\ref{fig:SC-optical}(c)) \cite{ref145}. Field-rotating MOKE and AHE measurements jointly indicate that the dominant contribution to the perpendicular magnetic anisotropy arises from interfacial effects at the Py/$\mathrm{RuO_{2}}$ boundary. The interface anisotropy energy is estimated to be approximately $0.77$ $\mathrm{erg/cm^2}$, highlighting the potential of $\mathrm{RuO_{2}}$ for spintronic interface engineering. Notably, this study did not observe the hallmark signature of spin splitting, namely, field-free magnetization switching. A plausible explanation is that the $\mathrm{RuO_{2}}$ thin films may host multiple $180^{\circ}$ antiferromagnetic domains with opposite Néel vector orientations. If these domains are uniformly distributed, the SST-induced spin currents generated in different domains, each carrying opposite spin polarizations, would mutually cancel. As a result, the net spin current vanishes, preventing the emergence of field-free switching.

Additionally, $\mathrm{RuO_{2}}$ exhibits a Faraday effect even in the absence of an external magnetic field, a consequence of time-reversal symmetry breaking inherent to its altermagnetic order \cite{ref146}. The analytically derived Faraday angle $\theta_F$ reaches the order of $10^{-5}$ rad at photon energies below the direct band gap, an order of magnitude larger than the Kerr angles observed in cuprate systems. This further underscores $\mathrm{RuO_{2}}$'s capacity to manifest robust magneto-optical responses despite the absence of net magnetization.

Recent symmetry-sensitive optical second-harmonic generation (SHG) and magneto-optical measurements have revealed a clear symmetry-breaking transition in ultrathin, fully strained $\mathrm{RuO_{2}}$ films. In films thinner than 4 nm, the SHG response undergoes a marked change near $\sim 500$ K. As shown in Fig.~\ref{fig:SC-optical}(d) , the SHG polar patterns display pronounced asymmetric indentations below 500 K (340 K and 420 K), whereas the high-temperature (540 K) response remains symmetric. This emergent asymmetry reflects the breaking of mirror symmetries preserved in the high-symmetry, nonmagnetic phase, thereby providing direct optical evidence of time-reversal symmetry breaking in the strained ultrathin limit. Notably, such SHG signatures are absent in strain-relaxed thicker films and bulk crystals, highlighting the critical role of epitaxial strain in stabilizing the magnetic phase \cite{ref70}.

%However, while thin-film transport and these magneto-optical studies provide compelling evidence for altermagnetism, bulk-sensitive measurements present a more conventional picture. Optical conductivity data (Fig. \ref{fig:SC-optical}e), obtained from Kramers-Kr\"onig transformations of reflectance spectra, closely match nonmagnetic band structure calculations, suggesting that the bulk electronic structure lacks clear signatures of altermagnetic order \cite{ref49}. Both experimental and computed optical conductivities display a sharp increase near a photon energy of $0.1$ eV and a pronounced minimum around $2$ eV, reinforcing the nonmagnetic model for bulk $\mathrm{RuO_{2}}$. This suggests that previously reported altermagnetic signatures may arise from strain or non-stoichiometry effects in thin films rather than intrinsic bulk properties \cite{ref36,ref70}.

\section{Outlook}

% Notes: this table can be further improved by adding information of film orientation (more detals of the sample) and temperature for each row.
\begin{table*}
\renewcommand\arraystretch{1.15}
\caption{Comparison of experimental measurements for ground magnetic state of $\mathrm{RuO_{2}}$. Abbreviations in the table: (spin-resolved) angle-resolved photoemission spectroscopy ((S)ARPES), anomalous Hall effect (AHE), anisotropic magnetoresistance (AMR), inverse altermagnetic spin-splitting effect (IASSE), magnetic circular dichroism (MCD), magnetoresistance (MR), neutron diffraction (ND), polarized ND (PND), planar Hall effect (PHE), quantum oscillations (QO), resonant X-ray scattering (RXS), residual resistivity ratios (RRR),  spin density wave (SDW), spin splitting torque (SST), spin-torque ferromagnetic resonance (ST-FMR), spin Seebeck effect (SSE), spin Hall magnetoresistance (SMR), scanning tunneling microscopy (STM), symmetry-sensitive optical second-harmonic
generation (SHG), tunneling anisotropic magnetoresistance (TAMR), time-reversal symmetry (TRS), X-Ray Magnetic Linear Dichro-
ism (XMLD), and muon spin rotation ($\mu$SR).}
\label{tab:Comparison table}

\begin{ruledtabular}

\begin{tabular}{ccccc}
\makecell{Magnetic \\ ground state} &  Experimental evidence (Method) & Sample & Ref. \tabularnewline
\midrule
\multirow{21}{*}{\makecell{Compensated \\
long-range\\ magnetic order}} 
 & $0.05$ $\mu_B$ magnetic moment (PND) & Bulk & \cite{ref4} \tabularnewline
 & Sharp antiferromagnetic diffraction (RXS) & Bulk and film (25 nm) & \cite{ref51} \tabularnewline
 & Antiferromagnetic diffraction signature (RXS) & Film (21 nm and 18.6 nm)  & \cite{ref14} \tabularnewline
 & Strain-induced $d$-wave spin polarization ((S)ARPES) & Ultrathin film (2.7 nm) & \cite{zhang2025observation} \tabularnewline
 & Spin splitting / $d$-wave spin polarization ((S)ARPES) & Bulk & \cite{ref28} \tabularnewline
 & TRS-breaking band structure (MCD)   &  Film (34 nm) & \cite{ref10} \tabularnewline
 & Néel vector parallel to the $[001]$ axis (XMLD) & Film (15 nm) & \cite{ref9} \tabularnewline
 & Large AHE with saturation beyond 50 T & Film (27/9.8 nm) & \cite{ref11,ref45} \tabularnewline
 & Magnetic response measured by AHE and PHE & \makecell{Film \\ (40nm, RRR $\sim$ 4.3)} & \cite{ref71} \tabularnewline
 & Strain stabilized magnetic ground state (AHE) & Ultrathin film (0.4 - 9.1 nm) & \cite{jeong2025metallicity} \tabularnewline
 & Angle-dependent XMLD/ spin-splitting MR (XMLD-MR) &  Film (30 nm) & 
 \cite{he2025evidence} \tabularnewline
 & Néel-vector orientation / strong temperature-dependent MR & Film (3 nm) & \cite{ref198} \tabularnewline
 & AM transition $<500$ K in ultrathin strained films (SHG)  & Ultrathin film ($\leq$ 4 nm) & \cite{ref70} \tabularnewline
 & Twofold angle-dependent third-order nonlinear Hall effect & Film (5 nm) & \cite{Chu2025nonlinear} \tabularnewline
 & Incommensurate SDW (ST-FMR) & Film (12 nm) & \cite{feng2024incommensurate} \tabularnewline
 & Néel-vector dependent SST (ST-FMR) & Film (6/12/10/15/10 nm) & \cite{ref5,ref3,ref21,ref15,ref200} \tabularnewline
 & ASSE-driven spin–charge conversion (SSE) & Film (12 nm/16 nm) & \cite{ref2,fan2024robust} \tabularnewline
 & IASSE signature in $\mathrm{RuO_{2}}(100)$/Py (THz emssion) & Film (5 nm)  & \cite{ref100} \tabularnewline
 & IASSE signature (SSE) & Film (20 nm/11.7 nm and 9.4 nm) & \cite{ref26,jung2025reversible} \tabularnewline
 & ASSE-driven large spin Hall angle / long spin diffusion length & Film (3-50 nm) & \cite{ref75} \tabularnewline
 & ASSE-driven $100\%$ field-free SOT switching & Film (15 nm) & \cite{ref110} \tabularnewline
 & ASSE-driven large field-like torque & Film (4 nm) & \cite{ref143} \tabularnewline
 & Giant spin-flop transition (TAMR) & Film (10 nm and 20 nm) & \cite{xu2025giant} \tabularnewline
 & Néel-vector-related magneto-transport properties (MR, AMR, SMR) & Film ($\sim$ 30.63 nm) & \cite{he2025magneticfield} \tabularnewline
\midrule 
\multirow{14}{*}{\makecell{Lack of \\ long-range\\ magnetic order} } & No spin precession signal ($\mu$SR) & Bulk (RRR $\sim$ 1500 ) &\cite{ref17} \tabularnewline
 & Magnetic order $<10^{-4}$ $\mu_B$ ($\mu$SR and ND) & \makecell{Bulk and film \\(11 nm/30 nm and 33 nm)} & \cite{ref22,akashdeep2025surfacelocalized} \tabularnewline
 & Magnetic order $<0.01$ $\mu_B$ (PND) & Bulk & \cite{ref8} \tabularnewline
 & Nonmagnetic electronic structure ((S)ARPES) & \makecell{Bulk and film \\ (13.5 nm/5 nm and 30 nm)} & \cite{ref27,ref210} \tabularnewline
 & \makecell{No magnetic or structural transition \\ (conductance and susceptibility)} & Bulk (RRR $\sim$ 158/152) & \cite{ref8,ref208} \tabularnewline
 & Nonmagnetic electronic structure (Optical conductivity) & Bulk (RRR $\sim$ 118) & \cite{ref49} \tabularnewline
 & \makecell{No magnetic phase transition up to 970 K \\ (Electrical and thermal transport)} & \makecell{Bulk \\ (RRR $\sim$ 80 and 12)} & \cite{ref60} \tabularnewline
 & No magnetic order at low temperature (DC susceptibility) & Bulk (RRR $\sim$ 1200) & \cite{paul2025growth} \tabularnewline
 & \makecell{No magnon mode in $\mathrm{RuO_{2}}$ films\\(Magneto-Raman spectroscopy)} & Film (32 - 36 nm) & \cite{abel2025probing} \tabularnewline
 & No SDW and surface magnetic instability (STM) & Ultrathin film ($<$ 3 nm) & \cite{kessler2025moireassisted} \tabularnewline
 & Nonmagnetic electronic structure (QO) & Bulk (RRR $\sim$ 400/250) &\cite{Wu_PRX_2025,qian2025determining} \tabularnewline
 & Crystal symmetry-independent MR & Film (5 nm) & \cite{akashdeep2025interfacegenerated} \tabularnewline
 & No IASSE signature of altermagnetism (SSE/ST-FMR) & Film (3.9 - 31.6 nm/12 nm) &\cite{wangrobust,ref118} \tabularnewline
 & No IASSE signature of altermagnetism (THz emission) & Film (12 and 5 nm/10 nm) &\cite{ref199,jechumtalspintochargecurrent} \tabularnewline
\end{tabular}

\end{ruledtabular}
\end{table*}

The recent surge of research interest in $\mathrm{RuO_{2}}$ originates from two developments. The first is the initial report of magnetism in $\mathrm{RuO_{2}}$, represented by the neutron diffraction measurement \cite{ref4}. The second is the theoretical proposal that $\mathrm{RuO_{2}}$ satisfies the symmetry requirements for altermagnetism, as demonstrated by the work \cite{ref40}. Subsequent debates regarding the existence and origin of magnetism have further intensified research activity. The debate surrounding the existence of altermagnetism in $\mathrm{RuO_{2}}$ still continues in current research. Table~\ref{tab:Comparison table} summarizes the key conflicting experimental measurements regarding the magnetism of $\mathrm{RuO_{2}}$. Given the numerous controversial results from experimental measurements, it is premature to draw a definitive conclusion on the existence of altermagnetism in $\mathrm{RuO_{2}}$. However, a closer examination of Table \ref{tab:Comparison table} reveals that experimental evidences supporting altermagnetism primarily come from thin-film samples, whereas those indicating a non-magnetic ground state are mainly from bulk samples. For pristine bulk $\mathrm{RuO_{2}}$, the community largely agrees on a non-magnetic ground state, supported by recent neutron diffraction, $\mu$SR, and ARPES studies, together with several indirect indications from bulk transport measurements. By contrast, ARPES signatures interpreted as evidence for altermagnetism remain isolated and are accompanied by only weak magnetic signals. Additional bulk-sensitive probes, particularly resonant x-ray scattering (RXS), are needed to further clarify the magnetic nature of pristine $\mathrm{RuO_{2}}$, as such measurements are currently scarce. For $\mathrm{RuO_{2}}$ thin films, many reports claim altermagnetic behavior, although most are based on indirect transport evidence. The magnitude of the ordered moment remains unknown, and there is no clear observation of a magnetic phase transition near the expected Néel temperature of 300–400 K. Recent reports showing the absence of IASSE in $\mathrm{RuO_{2}}$ films, despite earlier studies suggesting its presence, further highlight the strong sample dependence of thin film properties. The observed spin-torque signals can be accounted for entirely by the spin Hall effect. Consequently, it is premature to conclude that all $\mathrm{RuO_{2}}$ films are non-altermagnetic, given the substantial variability among samples. A key priority for future research is the direct detection of magnetic order in thin film samples that exhibit transport signatures consistent with altermagnetism. Our literature survey suggests that the ongoing controversies surrounding altermagnetism in $\mathrm{RuO_{2}}$ likely arise from four factors. First, most bulk-sensitive probes detect no magnetic order, while thin films often affected by epitaxial strain, stoichiometric deviations, or interfacial symmetry breaking, then exhibit magnetic signatures. Second, the magnetic response of $\mathrm{RuO_{2}}$ is highly sensitive to strain, Ru vacancies, and disorder, which may stabilize strain- or defect-induced magnetic states rather than intrinsic altermagnetism. Third, theoretical studies indicate that $\mathrm{RuO_{2}}$ resides near a magnetic quantum critical boundary, implying that any magnetic order is fragile and strongly sample-dependent, and may be tuned by small perturbations such as strain, defects, or doping. Fourth, several transport signatures frequently attributed to altermagnetism, including the anomalous Hall effect, spin-splitting torque, and tunnelling magnetoresistance, can also arise from impurities, interfaces, or surface magnetism, complicating their interpretation.

Techniques such as polarized neutron diffraction, which have a large penetration depth, still require sufficiently thick samples so that the signal is not dominated by the substrate.  In contrast, muon based methods include both conventional muon spin rotation for bulk studies and low energy muon spectroscopy for thin-film measurements. Because low energy muon spectroscopy can selectively probe near surface regions, it is well suited for detecting magnetic order in thin films. This distinction in measurement depth helps explain why thin-film and bulk studies sometimes lead to different conclusions regarding the magnetic ground state of  $\mathrm{RuO_{2}}$. Overall, the reported discrepant results suggest that slight differences in structural properties, such as strain or defects, between single-crystal thin films and bulk $\mathrm{RuO_{2}}$ may lead to significant differences in magnetic order. This is not rare in materials, where the reduced dimensionality and interface effects can fundamentally alter a material's ground state. Recent efforts have initiated the refinement of synthesis for defect-free, stoichiometrically controlled single crystals and thin films, concurrently exploring the impact of defects and strain on the magnetism. It is worth to consider that if the altermagnetism in $\mathrm{RuO_{2}}$ has sample dependency, a systematic characterization of the magnetic state through the integration of multiple high-sensitivity measurements (such as $\mu$SR, polarized neutron diffraction, and XMLD) can provide more compelling evidence and potentially reconcile the current discrepancies. 

Comparative insights from other altermagnetic materials can offer valuable guidance for advancing the study of $\mathrm{RuO_{2}}$. For example, the distinct altermagnetic order in MnTe has been revealed by advanced spatially resolved spectroscopic and imaging techniques~\cite{ref149}. The application of nanoscale X-ray magnetic circular dichroism (XMCD) combined with photoemission electron microscopy (PEEM) offers a direct mean to visualize the magnetic domains in $\mathrm{RuO_{2}}$. This approach is particularly useful to resolve the suspicion that the spin splitting effect in $\mathrm{RuO_{2}}$ may arise from surface- or defect-induced effects. Moreover, ARPES studies on MnTe have revealed both spin-degenerate altermagnetism and pronounced anisotropic band splittings \cite{ref155,ref157}. Temperature-dependent ARPES further tracked the evolution of these features across the magnetic transition and connected them to disordered local moment behavior \cite{ref157}, providing a framework for interpreting similar observations in $\mathrm{RuO_{2}}$. In addition, time-resolved magneto-optical Kerr effect (TR-MOKE) measurements in MnTe \cite{ref151}, capable of probing ultrafast spin dynamics and magnon excitations, suggest another promising direction. Extending such magneto-optical studies to the time domain in $\mathrm{RuO_{2}}$, and comparing the results with those from other altermagnets could reveal unique dynamic signatures of its magnetic ordering and clarify the various contributions to its optical response.

Despite lack of a definitive consensus on the altermagnetic order in $\mathrm{RuO_{2}}$, experimental observations of spin torque in thin films of the material are a promising sign for future applications. For example, the spin torque in $\mathrm{RuO_{2}}$ could be leveraged for use in magnetic random-access memory (MRAM). In these devices, spin torques driven by an electrical current enable efficient data writing by switching the magnetization of a storage layer. The SST in $\mathrm{RuO_{2}}$ is able to overcome the limitations of conventional STT and SOT in existing MRAM devices, which could lead to more efficient memory writing and better device performance. This makes further research into this direction especially worthwhile.

Meanwhile, theoretical studies suggest that the ongoing controversy surrounding the altermagnetism in $\mathrm{RuO_{2}}$ likely stems from its proximity to a quantum phase transition \cite{wu2007fermi, WuCJ2025} or from varying disorder levels~\cite{CALi25arXiv}. While current research primarily focuses on the identification of altermagnetism in $\mathrm{RuO_{2}}$, the manipulation of its altermagnetic order remains relatively unexplored. In principle, any means that potentially affect the electronic interactions in the $\mathrm{RuO_{2}}$ 4d orbitals can serve as a control parameter. By changing the state of the Fermi surface instability, the itinerant magnetism can be turned on and off. Therefore, a critical direction for exploring altermagnetism in $\mathrm{RuO_{2}}$ is finding an easily implementable control method to act as the tuning parameter. As a material positioned near a quantum phase transition point, $\mathrm{RuO_{2}}$ serves as an ideal platform for future studies dedicated to the control and tunability of altermagnetism.

%effects from disorder/impurities?

%review the research on the chirality splitting in magnons?

\section*{Conflict of interest}
The authors declare that they have no conflict of interest.

\begin{acknowledgments}
We acknowledge useful discussions with Xiong He. This work was supported by the National Key R\&D Program of China (2022YFA1403700), the National Natural Science Foundation of China (12525401, 12350402, 12488101,  12274258, and 12504059), Guangdong Basic and Applied Basic Research Foundation (2023B0303000011), Guangdong Provincial Quantum Science Strategic Initiative (GDZX2201001 and GDZX2401001), the Science, Technology and Innovation Commission of Shenzhen Municipality (ZDSYS20190902092905285), High-level Special Funds (G03050K004), the New Cornerstone Science Foundation through the XPLORER PRIZE, the start-up fund at HFNL, the Innovation Program for Quantum Science and Technology (2021ZD0302801), and Hubei Provincial Natural Science Foundation of China (2024AFB289).
\end{acknowledgments}

\section*{Author contributions}
Yu-Xin Li, Shuai Li and Song-Bo Zhang did the literature research for the article. Yu-Xin Li, Shuai Li, Song-Bo Zhang and Yiyuan Chen wrote the article. Hai-Zhou Lu and Shuai Li conceived and supervised the project. All authors contributed substantially to discussion of the content. 

\bibliography{reference}
\end{document}